\documentclass{aa}  
\usepackage{graphicx}

\usepackage{enumitem}
\usepackage{natbib}
\bibpunct{(}{)}{;}{a}{}{,}
\usepackage{txfonts}
\usepackage[normalem]{ulem}
\usepackage{color, soul}
\usepackage[breakable]{tcolorbox}

\begin{document} 

   \title{Modelling the evolution of the Galactic disc scale height traced by open clusters}

   \subtitle{}

   \author{Sandro Moreira\inst{1,2}       
        \and
        André Moitinho\inst{1,2}
        \and
        André Silva\inst{1,2}
        \and
        Duarte Almeida\inst{1} 
          }
    
   \institute{CENTRA, Faculdade de Ci\^{e}ncias, Universidade de Lisboa, Ed. C8, Campo Grande, P-1749-016 Lisboa, Portugal \and Laborat\'{o}rio de Instrumenta\c{c}\~{a}o e F\'{\i}sica Experimental de Part\'{\i}culas (LIP), Av. Prof. Gama Pinto 2, 1649-003 Lisboa, Portugal\\
              \email{sandro@sim.ul.pt, andre@sim.ul.pt, duarte.almeida@sim.ul.pt}
            }

   \date{Received 15 April 2024 / Accepted 22 December 2024}
  
  \abstract
    {The scale height (SH) of the spatial distribution of open clusters (OCs) in the Milky Way exhibits a well-known increase with age that is usually interpreted as evidence for dynamical heating of the disc or for the disc having been thicker in the past.}{We address the increase in the SH with age of the OC population from a different angle. We propose that the apparent thickening of the disc can be largely explained as a consequence of a stronger disruption of OCs near the Galactic plane by encounters with giant molecular clouds (GMCs).}{We present a computational model that forms OCs with initial masses and follows their orbits, while subjecting them to different disruption mechanisms. To set up the model and infer its parameters, we used and analysed a \textit{Gaia}-based OC catalogue. We investigate both the spatial and age distributions of the OC population and discuss the sample completeness. The simulation results are compared to the observations.}{ Consistent with previous studies, the observations reveal that the SH of the spatial distribution of OCs increases with age. We find that it is likely that the OC sample is incomplete even for the solar neighbourhood. The simulations successfully reproduce the SH evolution and the total number of OCs that survive with age up to 1 Gyr. For older OCs, the model-predicted SH starts deviating from observations, although it remains within the uncertainties of the observations. This can be related to the effects of incompleteness and/or simplifications in the model.
    }{The OC encounters with GMCs effectively explain the SH evolution of the OC population. An interesting result is that the average time for an object with a Sun-like orbit to encounter a GMC is approximately 700 Myr, aligning well with previous estimates for the Sun obtained through different methods.}


   \maketitle

\section{Introduction\label{sec:Introduction}}

Understanding the evolution of galaxies involves tracking their morphological changes and identifying the mechanisms driving these transformations.
The Milky Way (MW)'s thin disc, which hosts most of the stars in the Galaxy, has been continuously forming stars for more than ten billion years. This activity has led to a complex history of stellar dynamics interwoven with interactions with the interstellar medium, shaping the disc throughout its existence.

To a good approximation, the vertical number density of stars in the Galactic disc decreases exponentially with distance from the disc's plane \citep{1980ApJS...44...73B}. The exponential decrease is characterised by the decay factor, referred to as the disc ‘scale height’ (SH), which parametrises the thickness of the disc.
Different populations of objects may display different vertical distributions. For instance, the distribution of stars can have a SH from $\sim 40 \pm 7$ pc at log(age) $\sim 7.5 - 7.7$ up to 211 $\pm 13$ pc at log(age) $\sim 9.3 - 9.5$ \citep{Mazzi_2024}. Although the thickness of the disc can be studied using various types of objects, such as stars of different spectral types, open clusters (OCs) are often used as disc probes due to the advantages that they provide from an observational standpoint \citep[e.g.][]{moitinho_observational_2009}. In particular, OC ages can be determined with an accuracy out of reach for most stars, which allows for the direct inspection of specific times in the evolution of the Galaxy. 

Because OCs are bound groups of stars with different masses immersed in the Galactic potential, a variety of both internal and external factors contribute to their evolution and, ultimately, dissolution. Thus, studying the evolution of the OC population, particularly its vertical distribution, is key to understand the underlying mechanisms that shape the structure of the MW and its evolution. 

The SH of the spatial distribution of OCs in the MW exhibits a well known increase with age and has been discussed multiple times in past studies \citep[e.g.][]{Lynga1982, Lynga1985, JanesPhelps1994, Bonatto2006, BucknerFroebrich, cantat-gaudin_painting_2020, Joshi_2023}. Table \ref{table:SH_studies} shows the SHs obtained for different ages in a non-exhaustive list of previous studies. The table, which includes SH determinations derived from data obtained by the \textit{Gaia} mission \citep{2016A&A...595A...1G, 2018A&A...616A...1G, 2021A&A...649A...1G} as well as from before \textit{Gaia}, clearly shows that the OC SH increase with age is a well-established observational result.

\begin{table}
\caption{SH of the OC distributions from previous studies.}
\label{table:SH_studies}
\centering
\begin{tabular}{ccc}
\hline\hline
Reference                &  Age (Myr)       & S$_\text{H}$ (pc)   \\ \hline
1       &   $< 100$              & 60                  \\
                         &  100 - 1000            & 80                  \\
                         &   $>$ 1000             & 200                 \\ \hline
2  &   Young                & 55                  \\
                         &    Old                 & 375                 \\ \hline
3      &   $< 200$              & 47.9 $\pm$ 2.8      \\
                         &  200 - 1000            & 149.8 $\pm$ 26.3    \\
                         &   $ > 1000$            & $\infty$            \\ \hline
4 &   1                    & 40                  \\
                         &   1000                 & 75                  \\ 
                         &   3500                 & 550                 \\ \hline
5          &   100                  & 74 $\pm$ 5      \\
                                             &   1000                 & 142 $\pm$ 7     \\ \hline
6          &   $< 20$               & 70.5 $\pm$ 2.3      \\
                         &  20 - 100              & 87.4 $\pm$ 3.6      \\ \hline
7       &   $< 20$               & 80.8 $\pm$ 8.0      \\
                         &  20 - 700              & 96.4 $\pm$ 1.9      \\
                         &  700 - 2000            & 294.7 $\pm$ 19.5    \\ \hline
\end{tabular}
\tablebib{
(1) \citet{Lynga1985} ; (2) \citet{JanesPhelps1994}; (3) \citet{Bonatto2006}; (4) \citet{BucknerFroebrich};
(5) \citet{cantat-gaudin_painting_2020}; (6) \citet{Hao2021}; (7) \citet{Joshi_2023}.
}
\end{table}

The increase in the SH with age of the OC population is usually interpreted in a way similar to that of the SH evolution of the stellar population: as evidence for the disc having been thicker in the past \citep{JanesPhelps1994}, or for dynamical heating of the disc due to scattering events, such as interactions with the spiral arms and giant molecular clouds \citep[GMCs, ][]{gustafsson_gravitational_2016, BucknerFroebrich}. 
However, as was pointed out by \citet{Friel_1995}, the encounters leading to such orbit perturbations could also disrupt the clusters themselves. Moreover, \citep{LamersGieles2006} showed that encounters with GMCs may be the main mechanism responsible for the disruption of OCs in the solar neighbourhood. On the kinematic side, \citet{Tarricq_2021} studied the age dependence of the velocity dispersion of the OC population, finding a much smaller vertical dynamical heating rate compared to that of stars. They suggest that, while interactions with GMCs are likely the primary cause of scattering for field stars, these interactions may be less efficient at scattering OCs, or alternatively, that OCs are destroyed during these interactions, and therefore do not contribute to an increase in velocity dispersion.
In any case, these arguments point out possible factors, but no actual astrophysical modelling has been published to reproduce the SH evolution of the OC distribution.

In this context, we follow the suggestion of \citet{moitinho_observational_2009} and address the increase in the SH of the OC population with age from a different angle: that the increase in the SH might be a consequence of a stronger disruption near the Galactic plane (GP) due to disc phenomena such as encounters with GMCs. The OCs formed at larger distances from the GP are likely to survive longer, because they not only spend most of their lifetimes at larger distances from the plane, but also cross it with higher velocities, having less time to be affected by the disruption mechanisms. As time passes, the distribution is gradually trimmed at lower heights, resulting in a flattening of the height distribution and an apparent increase in the SH with time. We present a dynamical model that follows the orbits of OCs and includes their disruption due to interactions with the disc and mass loss due to stellar evolution and evaporation. 

This paper is structured as follows. In Sect. 2, we present the observational sample of OCs and discuss statistical properties that are relevant for this study, such as the completeness of the sample, as well as the spatial and age distributions. In Sect. 3, we present the proposed model and how we set up the parameters related to the integration of the orbits, the initial masses of the OCs, and the mechanisms of disruption. Section 4 provides a more in-depth analysis of the parameter space in which we infer the free parameters of the model by grid-searching optimal solutions. In Sect. 5, we discuss the obtained results. Final considerations and conclusions of this study are presented in Sect. 6.

\section{Observational data\label{sec:ObservationalData}}

\subsection{Overview}

The high-precision astrometric and photometric data from \emph{Gaia} mission have allowed significant progress in building an observational base of OCs. They have brought about the discovery of hundreds of new OCs \citep[e.g.][]{Castro-Ginard, LiuPang, 2019A&A...624A.126C, sim_207_2019, Ferreira_2020, cantat-gaudin_painting_2020, Tristan_2018, Jaehnig_2021, 2023A&A...673A.114H}, the identification of their members, and the systematic determination of their fundamental parameters \citep[e.g.][]{Cantat2018, bossini_age_2019, carrera_open_2019, monteiro_distances_2019, cantat-gaudin_painting_2020, monteiro_fundamental_2020, 2023A&A...673A.114H}.

The OCs parameters used in this study are from the catalogue of \citet{DiasCat}, which uses the data from \emph{Gaia} Data Release 2 (DR2) and contains a total of 1743 clusters along with their ages, distances, proper motions, metallicities, and many other properties. The individual stellar memberships were collected from other studies \citep[full list in][]{DiasCat}. This catalogue provides homogeneous determinations of ages and distances, among other parameters, which were obtained using the same isochrone set and fitting method as is described in \citet{Monteiro2017} and \citet{monteiro_fundamental_2020}.

To study the dependence of the SH on age, we proceeded similarly to  \citet{JanesPhelps1994, Bonatto2006, Joshi_2023} and split the OCs sample into different age groups. Thanks to the larger number of OC parameters available for this study, we included an additional age group. Table \ref{agegroups} shows the selected age intervals for each group with the corresponding sample sizes.

\begin{table}
\caption{Age groups for the sample of OCs considered in this study.}
\label{agegroups}
\centering
\begin{tabular}{ccc}
\hline\hline
                       & Age (Myr) & Sample Size \\ \hline
Young              & 0 $<$ Age $\leq 200$    &  969  \\
Intermediate Young & 200 $<$ Age $\leq 500$  & 333 \\
Intermediate Old   & 500 $<$ Age $\leq 1000$ & 250 \\
Old                & Age $> 1000$ &  191 \\ \hline
All                    & - - & 1743  \\  \hline    
\end{tabular}
\end{table}

\subsection{Distribution on the galactic plane and completeness}

We now proceed to characterise our observational sample,  starting by considering the sample completeness.
The completeness of the OCs census has been discussed multiple times in the literature, with different studies claiming that the sample of OCs is complete in the solar neighbourhood up to distances of $\sim 1$ kpc \citep{KarchenkoCompleteness, Piskunov2006, BucknerFroebrich} from the Sun, and some even suggesting  greater distances  of $\sim$ 2-3 kpc \citep{Joshi_2005}, although those claims have been questioned by \citet{moitinho_observational_2009}. The recent avalanche of OC discoveries mentioned above has clearly shown that
assumptions of completeness may not be correct, even in the solar neighbourhood. 
To assess the dependence of completeness on distance, we split the OCs into sub-samples, considering three cylindrical cuts with radii of 1.00, 1.75, and 2.50 kpc.

\begin{figure*}
\centering
    \includegraphics[width=0.49\textwidth]{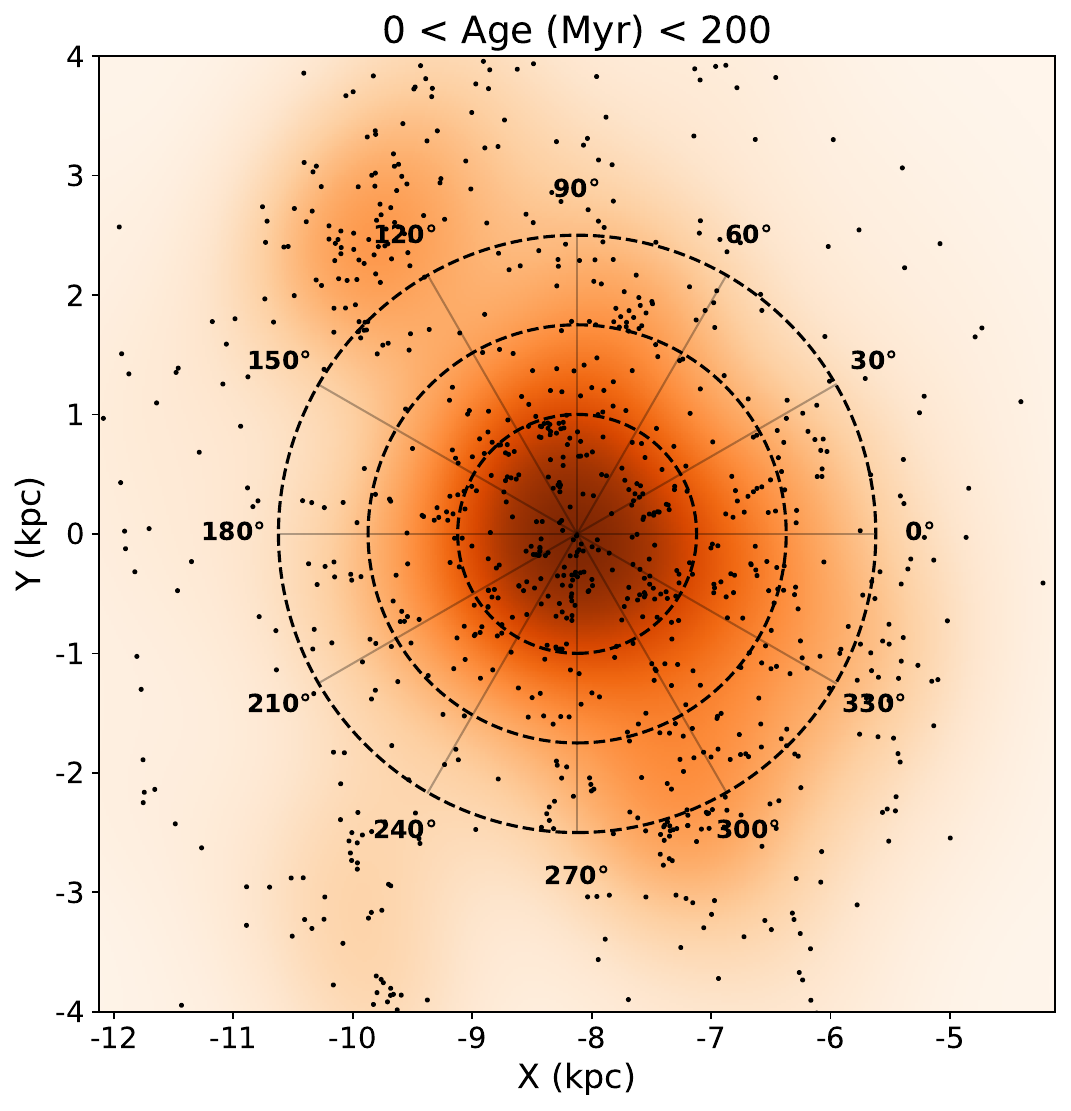}
    \includegraphics[width=0.49\textwidth]{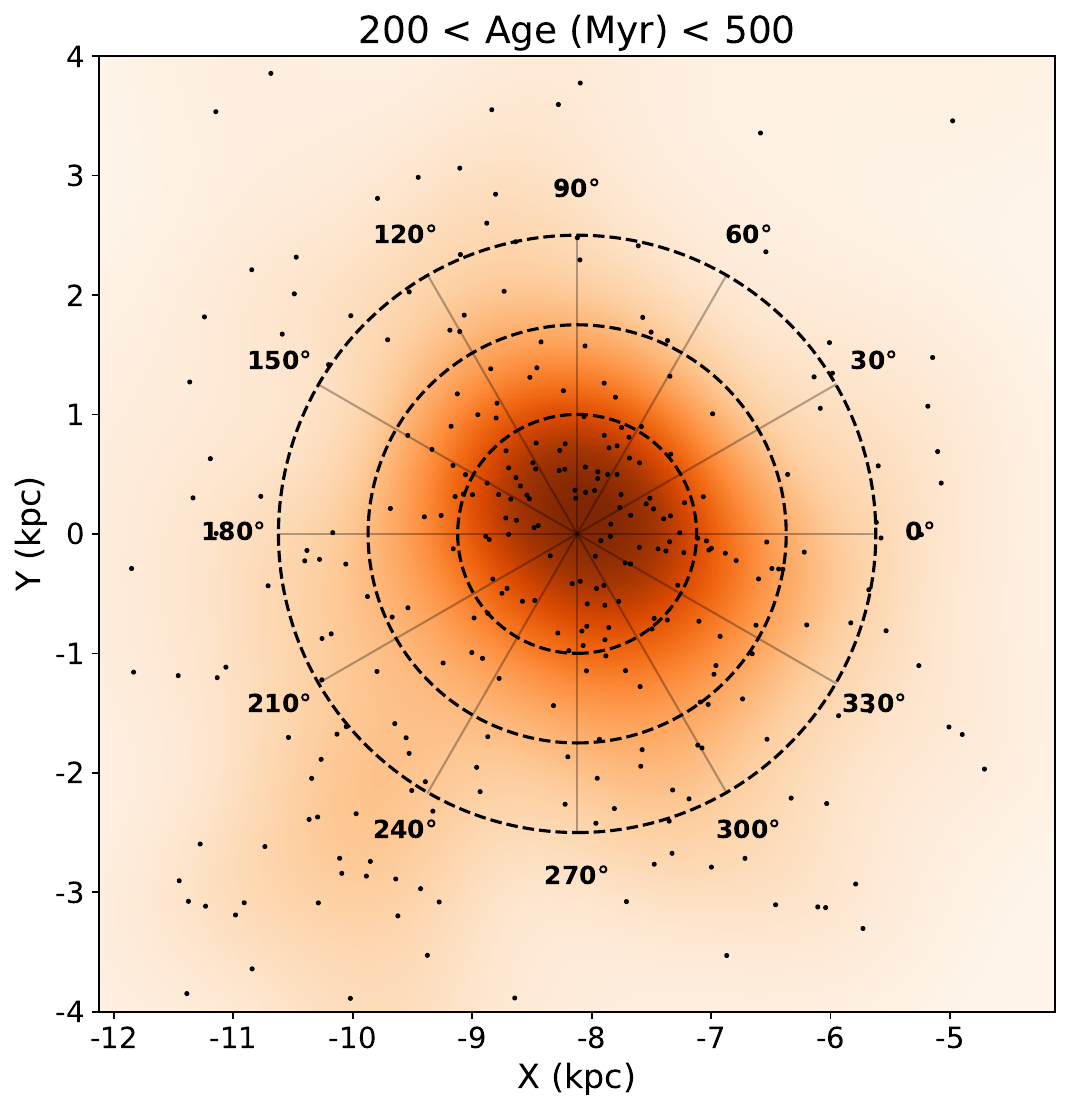}\\
    \includegraphics[width=0.49\textwidth]{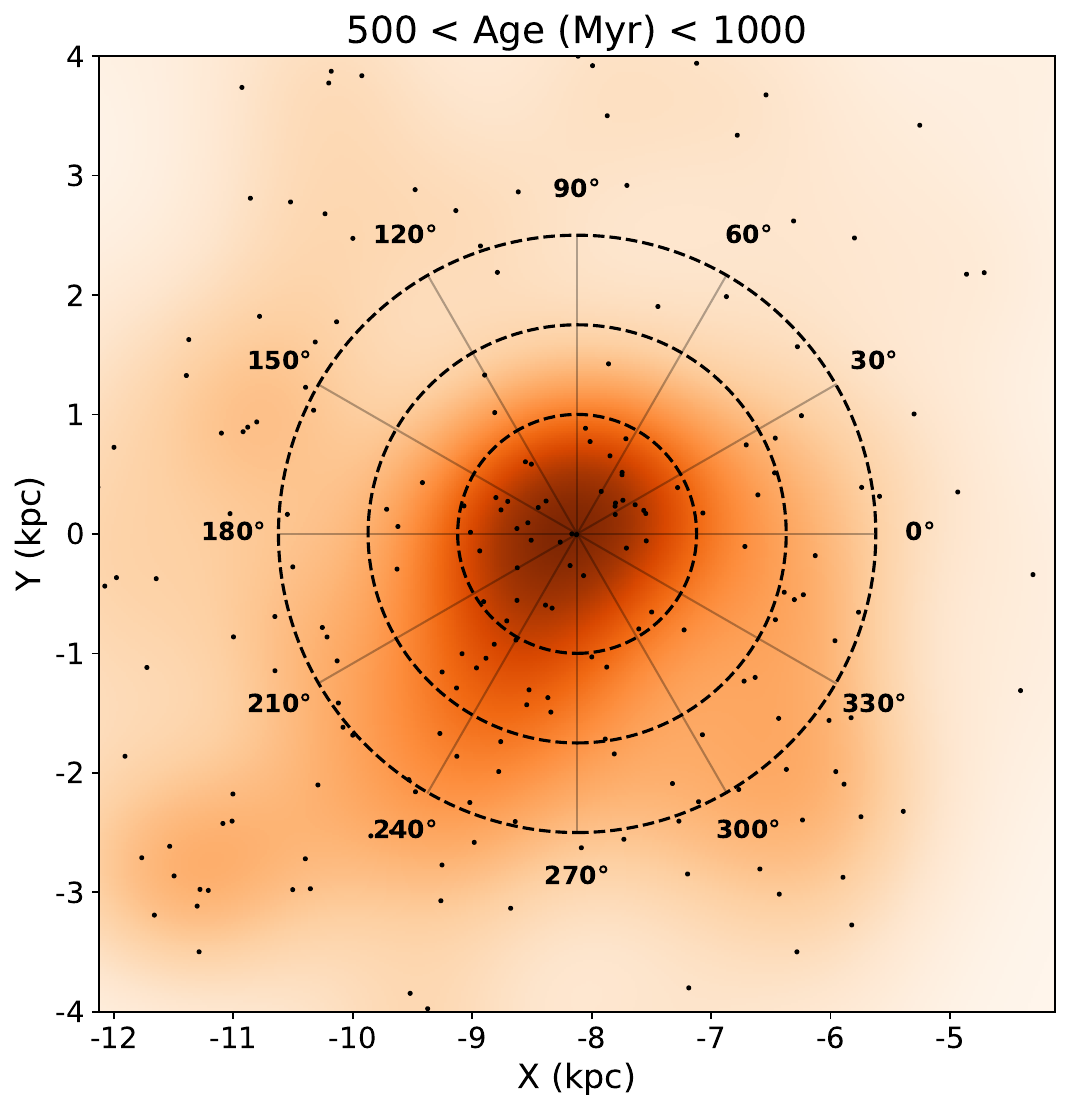}
    \includegraphics[width=0.49\textwidth]{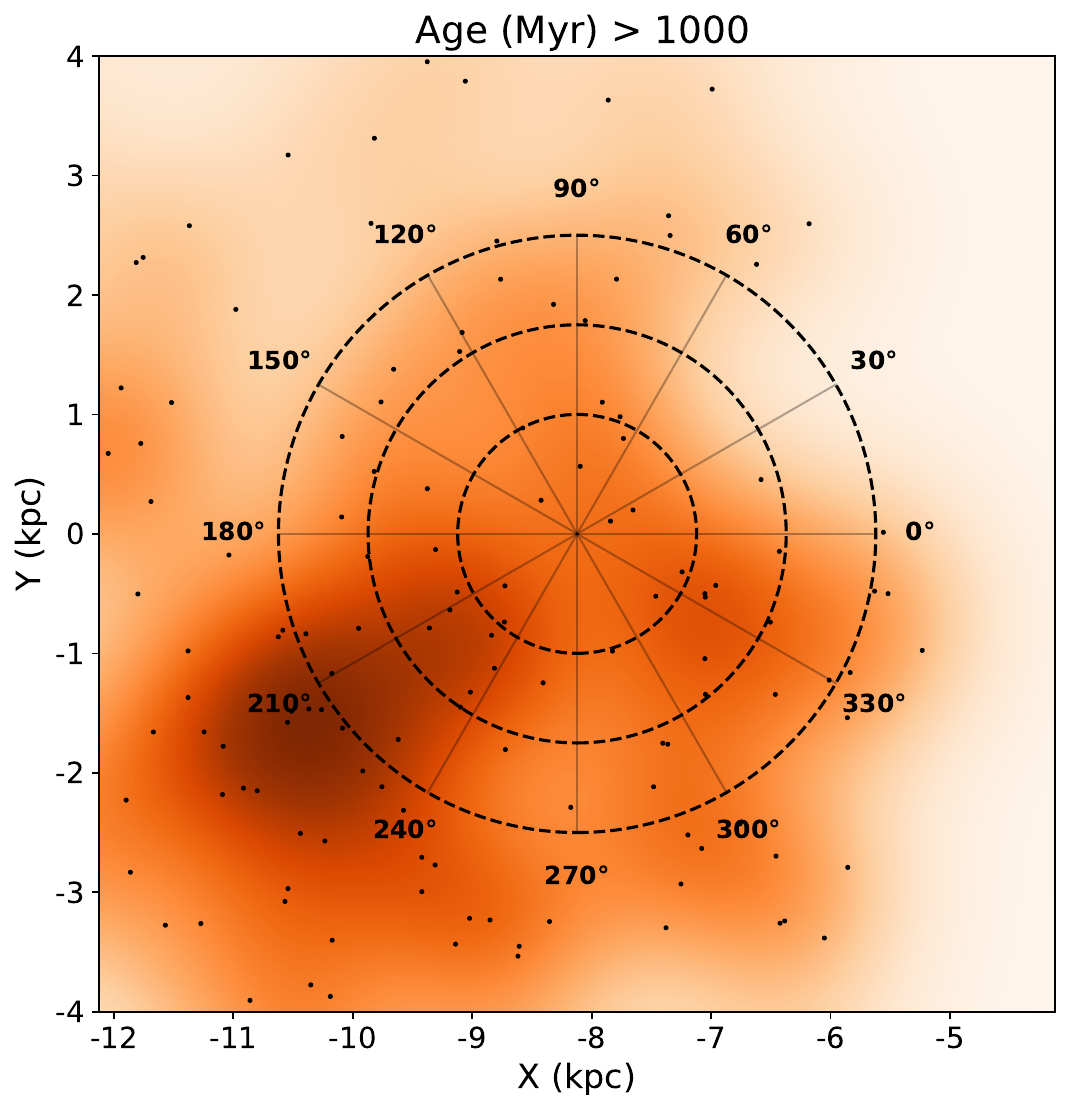}
    
    \caption{Distribution of OCs catalogued in \citet{DiasCat}, projected on the GP, in galactocentric co-ordinates, for the different age groups used in the study. The Sun is located at (-8.122, 0) kpc. The concentric circumferences represent the radii of the cylindrical cuts considered: 1.00, 1.75, and 2.50 kpc. The colour scale encodes the number density of OCs on the plane, with darker tones representing higher density.}
    \label{galacticPlane}
\end{figure*}

Figure \ref{galacticPlane} shows, for each age group, the projected spatial distribution of OCs on the GP using the galactocentric co-ordinate frame from \emph{astropy SkyCoord} \citep{Astropy}. The Sun is located at (X, Y) = (-8.122, 0) kpc, with the angular co-ordinate corresponding to the Galactic longitude. Dashed circumferences represent the cylindrical cuts considered. The plots also include kernel density estimations (KDEs) of the number density of OCs on the XY plane, using a Gaussian kernel with a bandwidth of 650 pc. The choice of a large bandwidth was made within the context of addressing incompleteness rather than emphasising local structures. The plots reveal that the sample of OCs does not exhibit an homogeneous spatial distribution on the plane. A systematic decrease in the number of observed OCs with the distance to the Sun is clearly seen. 

\begin{figure}
\centering
\resizebox{\hsize}{!}{\includegraphics{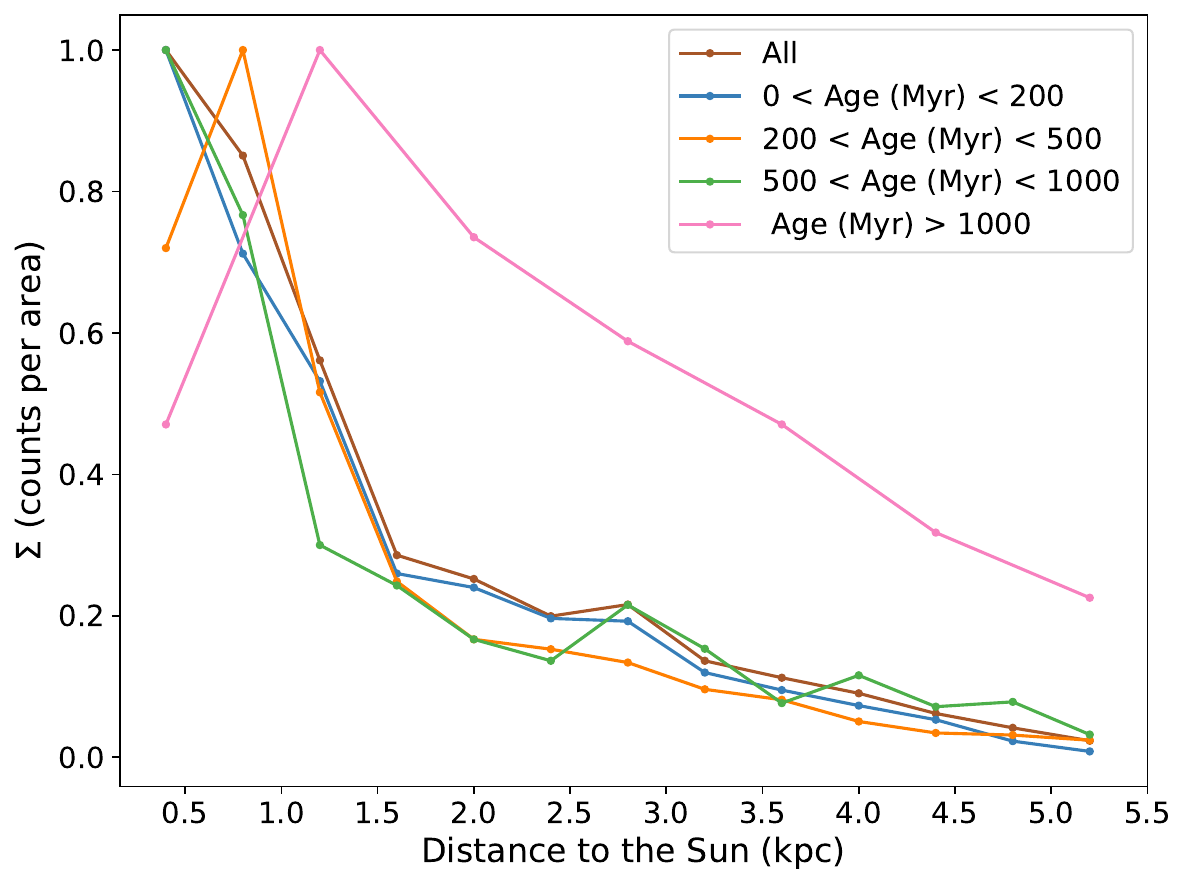}}
    \caption{{Surface number density of OCs vs heliocentric distance. The density was computed as the number of OCs in concentric cylindrical rings divided by the areas of the rings. The plotted values are scaled to a maximum value of 1 for comparison purposes. The small circles mark the limits of the rings.}}
    \label{surfacedensity}
\end{figure}

Assuming that our position in the Galaxy is not special, for a complete sample the surface density of the number of OCs should remain roughly constant in our neighbourhood. Figure \ref{surfacedensity} shows the surface density, in concentric rings, for different radial heliocentric distances and for the considered age groups. The initial bins correspond to circles with a radius of 0.4 kpc and subsequent rings then increase the radius by 0.4 kpc for age groups younger than 1000 Myr, and 0.8 kpc for the older age group. The curves are normalised so that the maximum surface density equals unity, for each age group.  With the exception of the old OCs, all age groups exhibit a similar decline in the surface density. The trends observed in other observational catalogues are similar to the ones displayed in age groups with ages $< 1000$ Myr \citep[see][Fig.\ 1]{BucknerFroebrich}. This is a strong indication that the sample of OCs is not complete for distances $< 1.5$ kpc or even $1$ kpc.

Surprisingly, the old age group displays a less accentuated  decline in the surface density with distance. Considering that older OCs are typically fainter, the opposite would be expected, especially when seen against crowded fields. We suggest that the selective destruction of OCs near the plane, incidentally the mechanism considered here for the SH evolution, could explain this. Most, if not all, of the older OCs that formed at closer distances to the GP may have already been disrupted. Thus, the ones that remain are located at greater distances from the GP, where reduced background contamination facilitates their detection, leading to the older group being more complete than the others. 

Ideally, we would employ complete samples in our analysis; however, selecting samples in such a small volume comes with the trade-off of having a reduced number of OCs, and thus worse statistics. Furthermore, local structures can also lead to a biased representation of the population of OCs, as we shall see in the following discussions. In the next sections, we attempt to find a balance among these trade-offs by exploring the properties of the different cylindrical cuts.

\subsection{Age distribution \label{sec:AgeDistribution}}

\begin{figure}
    \centering
    \resizebox{\hsize}{!}{\includegraphics{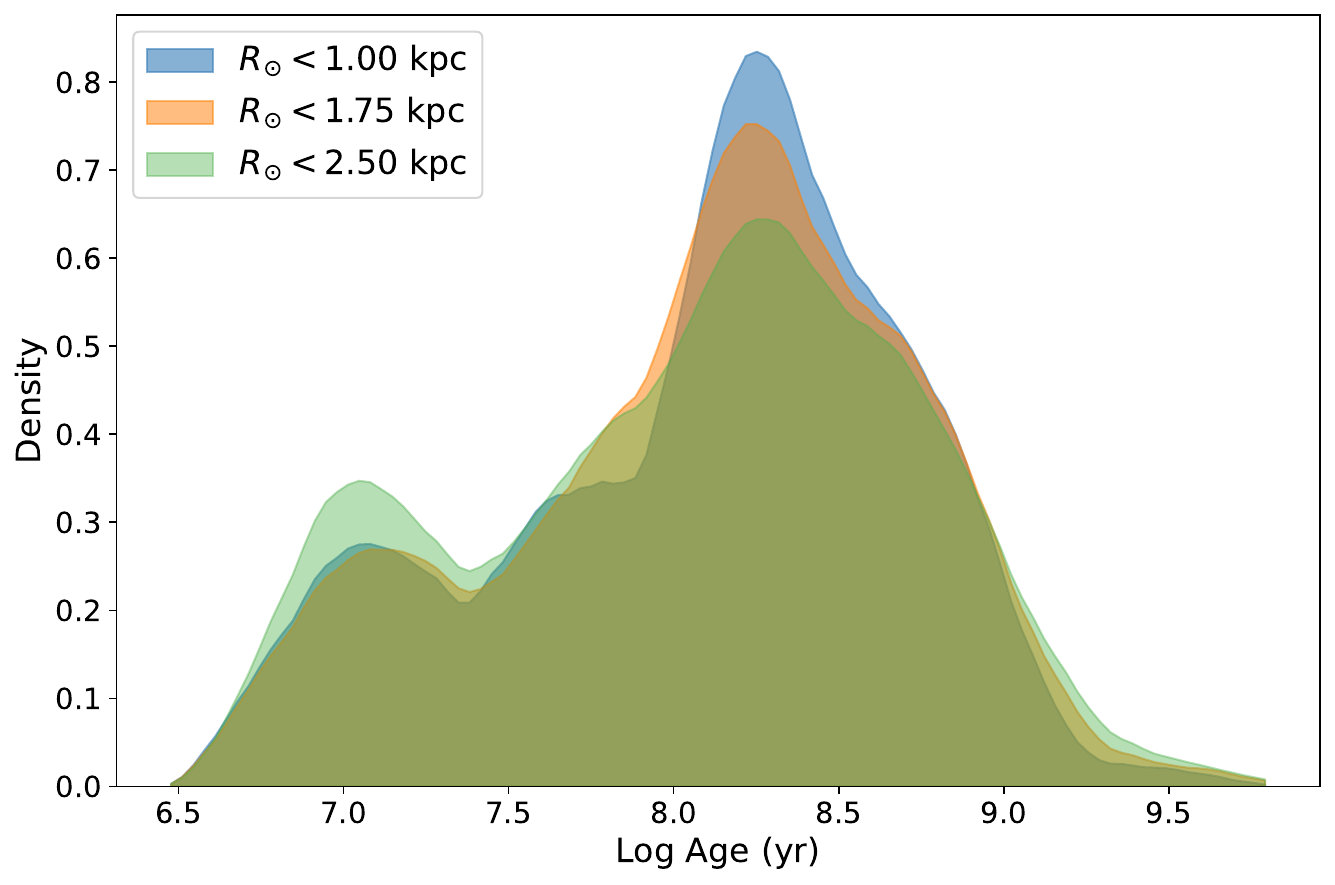}}
    \caption{KDEs for the age distribution of OCs in different cylindrical cuts. The blue KDE corresponds to R = 1.00 kpc, the orange one to R = 1.75 kpc, and the green one to R = 2.50 kpc.}
    \label{ageD}
\end{figure}

Figure \ref{ageD} shows KDEs for the age distribution of the OC population in the considered cylindrical cuts, using an Epanechnikov kernel \citep{Epanechnikov} with a bandwidth of log(age) = 0.25. The minimum age observed is log(age) $\sim$ 6.6. This corresponds to the expected age at which OCs leave the GMCs where they were formed \citep{Lada_2003}. Most younger clusters will only be visible at infrared wavelengths, and thus not observed by Gaia.

If there is no disruption, the number of OCs should increase with age. However, as is discussed in Sect. \ref{sec:Introduction},  OCs eventually dissolve into the field population. Hence, the KDEs display a peak around the log(age) $\sim 8.2$ ($\sim 150$ Myr) that reflects the typical timescales for OC disruption. The peak becomes less evident as the heliocentric distance increases. This can be attributed to a lower completeness at large radii, as is already seen in figures~\ref{galacticPlane} and~\ref{surfacedensity}. 

The number of OCs substantially decreases as we approach log(age) $\sim$9 ($\sim 1$ Gyr). It is expected that only OCs formed with very large masses are able to survive this long \citep{Lamers2005_B}. Consistent with the earlier discussion regarding the completeness of the OC population, the observed trend indicates that the fraction of old OCs is higher as we consider larger heliocentric distances. This trend supports the idea that the completeness of the older age group exhibits a distinct decrease compared to the other age groups.

Finally, we notice a bump around log(age) $\sim$ 7.0, mostly in the 2.5 kpc cut although it can be marginally identified in the closer cuts. The bump is then followed by a dip around log(age) $\sim$ 7.4, which seems to compensate for the excess in the bump. It is hard to assess the significance of these features, since they are seen in some (Gaia-based) catalogues \citep{DiasCat,cantat-gaudin_painting_2020} but not in others \citep{2023A&A...673A.114H, 2024AJ....167...12C}. Overall, the tendency is for the number of clusters to increase, with some fluctuation, until peaking at log(age) $\sim$ 8.2. This is an issue we shall investigate further in future work.

\subsection{Vertical distribution\label{sec:V_dist}}

\begin{figure*}
\centering
    \resizebox{0.33\hsize}{!}{\includegraphics{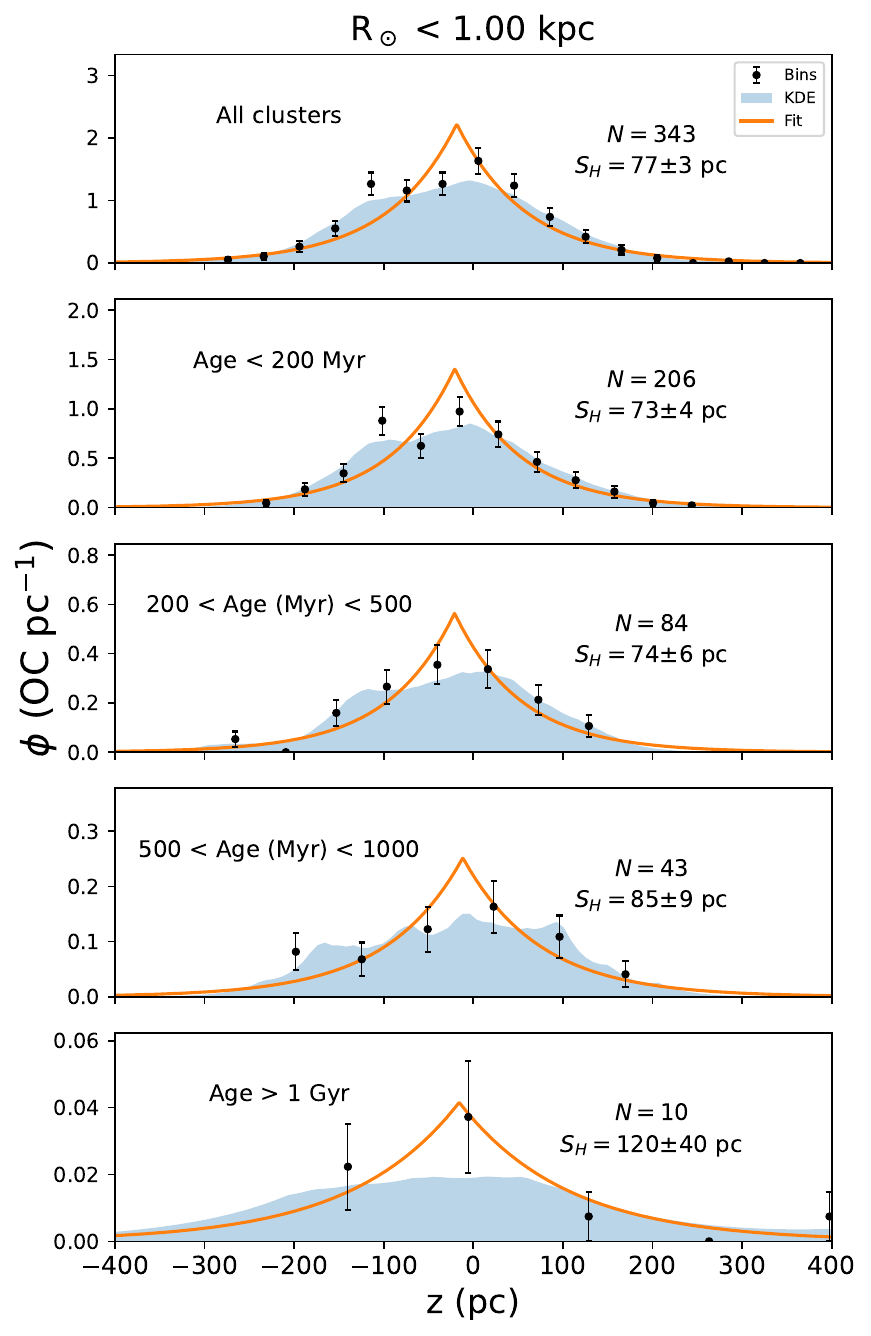}}
    \resizebox{0.33\hsize}{!}{\includegraphics{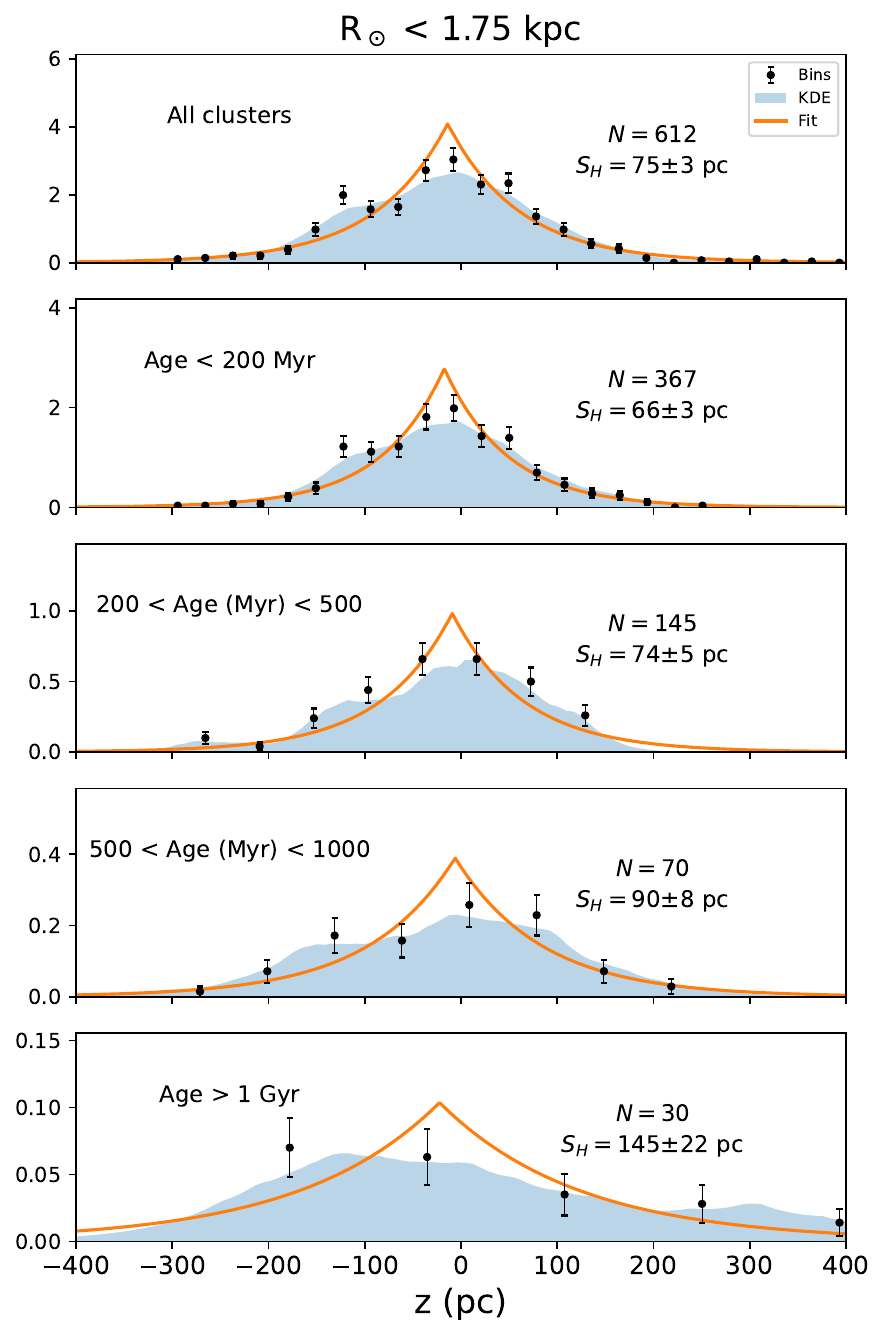}}
    \resizebox{0.33\hsize}{!}{\includegraphics{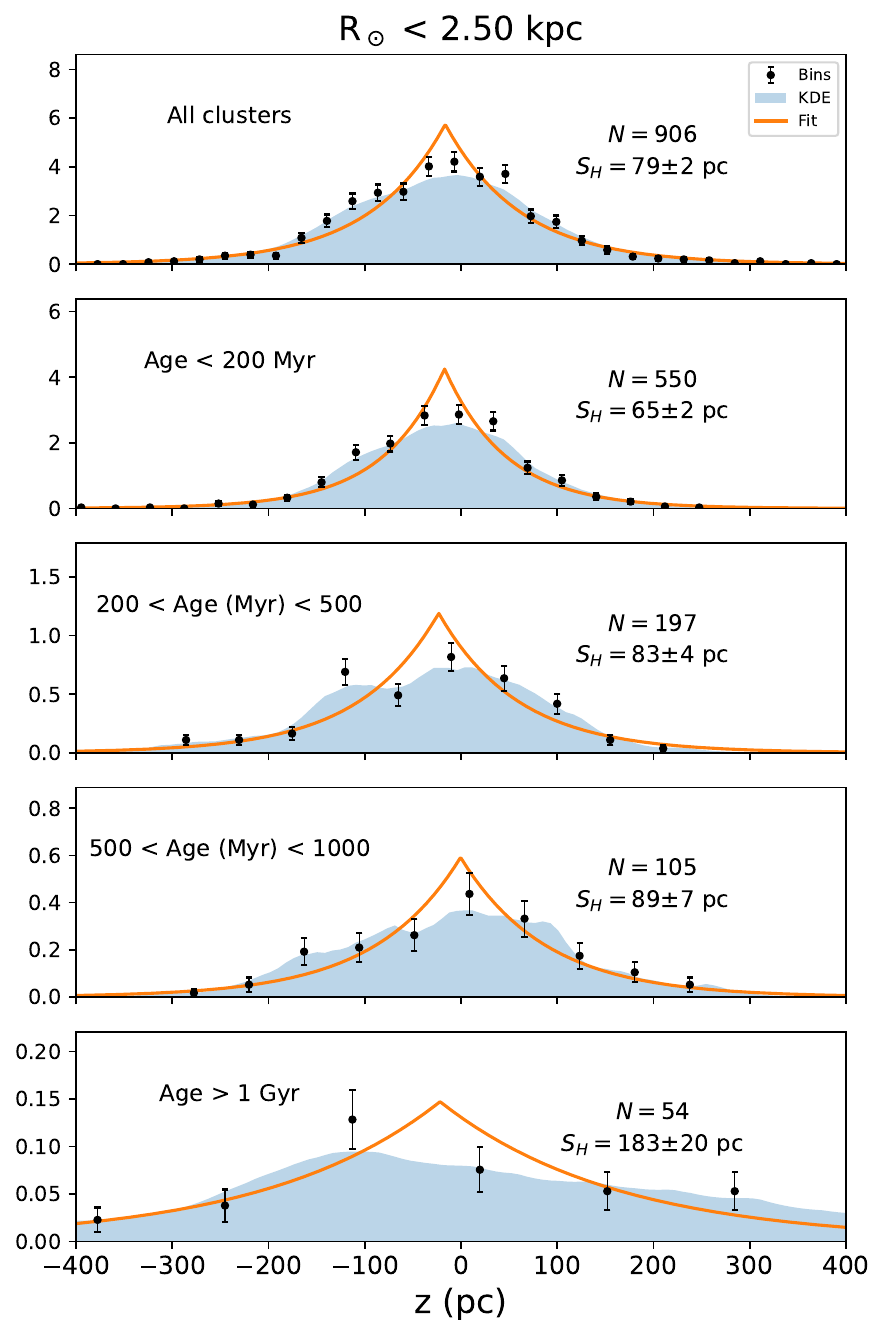}}
    \caption{Vertical distribution of the OCs catalogued in \citep{DiasCat} on the GP for different age groups and for the considered cylindrical cuts. The filled curve represents a KDE using an exponential kernel and the solid lines {represent maximum likelihood estimations of Laplace profiles (see Sect. \ref{sec:V_dist} for details)}.}
    \label{VerticalD}
\end{figure*}

The vertical distributions of OCs for the different age groups and cylindrical cuts are represented in figure~\ref{VerticalD}. The plots present, in blue, KDEs  of the distribution using an exponential kernel. The uncertainties of the bins correspond to a $1\sigma$ Poisson error and the widths of the bins were obtained using Knuth's rule \citep{KnuthRule}. The SHs were calculated analytically with its maximum likelihood estimator assuming a Laplacian (double exponential) distribution:
\begin{equation}
    \phi (Z; \text{ } S_H, Z_{\odot}) = \frac{1}{2 S_H}e^{- (\vert Z - Z_{\odot}\vert) / S_H}
    \label{eq:SH}
,\end{equation}
where $Z_{\odot}$ is the median of the distribution and a measure of the Sun's vertical displacement in relation to the GP. The SH maximum likelihood estimator can be computed analytically, being the mean absolute deviation from the median. The values presented for the SH are the average of the 1000 bootstrap runs \citep{EfronBootstrap} and the respective errors are the standard deviations. The orange lines represent the {corresponding} Laplacian curve.

Figure~\ref{VerticalD} readily shows that all cylindrical cuts present a SH increase with age, similarly to what was reported in previous studies (such as those listed in Table~\ref{table:SH_studies}). Nevertheless, we note that with respect to the older pre-Gaia studies, the increase in the SH with age is less accentuated; similarly to \citet{Joshi_2023} and in contrast to \citet{Bonatto2006}, we find a large, yet finite, SH for the older OCs. This is a clear result of using a more complete catalogue with more accurate determinations of OC properties.

For the age groups with ages $< 1$ Gyr, the SH evolution is consistent, within the errors, between the different cylinder cuts. This suggests that the effects of incompleteness do not impact, at least to a significant extent, the SH evolution. 
{We remark that this is not unexpected. As will be clear from the description of the SH  evolution model in Sect.~\ref{sec:model}, the evolution of the SH is determined by the relative numbers of OCs in different age groups (at a given position), not the absolute numbers. Thus, verifying that the numbers of OCs in the different age groups follow a similar distance-dependent pattern reassures us that our results are not significantly affected by incompleteness, as is seen in Fig.~\ref{VerticalD}}.

For the last age group, the large uncertainties make it challenging to draw any conclusions about the influence of the sample incompleteness. Nevertheless, the SH seems to increase by roughly 20 pc in subsequent cylindrical cuts.  It is conceivable that, whilst the decline of the completeness for the older age group is less pronounced compared to the other age groups, some missing OCs may be located closer to the GP, where their detection is more challenging. Given the relatively small sample size of the older age group, the absence of just a few OCs closer to the GP can significantly affect the estimated SH value.

Finally, it  worth noticing that the OCs with ages $< 200$ Myr for R$_\odot < 1$ kpc  have a relatively high SH in comparison to the other cylindrical cuts. It is possible to notice a peak in the vertical distribution of OCs at roughly $ -150 < \text{z (pc)} < -100$ for the young. This corresponds to the Orion  star-forming region. 

\subsection{Sample selection}

The 2.50 kpc cut is very likely to be significantly incomplete and the 1.00 kpc cut has a very reduced number of old OCs, which leads to large uncertainties. Moreover, the impact of local structures (Orion) on the vertical distribution is evident due to the reduced number of OCs in this cut. Thus, our final selection is the cylinder with a radius of 1.75 kpc, as we believe this selection finds a well-balanced compromise of the discussed trade-offs.

Furthermore, we restricted our analysis to OCs with ages up to of 2 Gyr. As is detailed in Sect. ~\ref{sec:AgeDistribution}, the population of OCs older than $\sim$ 1 Gyr is notably small. Only six OCs were removed, constituting 20\% of the old population and roughly 1\% of the total number of OCs in the 1.75 kpc cut. This will substantially decrease the computational time, {since we shall not be following the OC population up to the ages of the oldest catalogued objects $\sim 10$ Gyr,}. The SH of the old OCs (1 Gyr $<$ age $<$ 2 Gyr) group changes from $145 \pm 22$ pc to  $134 \pm 21$ pc and the SH of all OCs (age $<$ 2 Gyr) changes from $75 \pm 3$ pc to $74 \pm 2$ pc. Table \ref{table:OC_Properties} shows the parameters of interest for the selected cut. 

\begin{table}
\centering
\caption{Properties of the vertical distribution of the OC samples in the cylinder cut, $R_{\odot} \leq 1.75$ kpc.}
\label{table:OC_Properties}
\begin{tabular}{c c c c c}
\hline\hline
Age Group (Myr)    & Sample Size    & $S_H$ (pc)         & $Z_{\odot}$ (pc)   \\ \hline
$\leq 2000$        &  606           & 74 $\pm$ 2         &  14 $\pm$  3            \\
0 - 200            &  367           & 66 $\pm$ 3         &  17 $\pm$  4            \\
200 - 500          &  125           & 74 $\pm$ 5         &  9  $\pm$  7           \\
500 - 1000         &  90            & 90 $\pm$ 8         &  6  $\pm$  6           \\
1000 - 2000        &  24            & 134 $\pm$ 21       &  21 $\pm$  26           \\
$>$ 2000           &  6             & --                 &  --                 \\

\hline\
\end{tabular}
\end{table}

\section{The model\label{sec:model}}

\subsection{Overview}

As is discussed in Sect. \ref{sec:Introduction}, we propose that the apparent increase in the OC SH with age can be explained by the stronger disruption near the GP due to interactions with the GMCs. However, encounters with GMCs are not the only mechanism of disruption that the OCs are subjected to throughout their lifetimes. Internal processes, such as stellar and dynamical evolution, contribute to a gradual mass loss of the OC, and to their eventual dispersion. We refer to the mass loss by dynamical evolution as the mass lost due to both evaporation and the gravitational stripping effects of the Galactic tidal field. Accounting for these processes is essential to reproduce the total number of OCs that survive with age. The computational model thus has the following ingredients:

\begin{enumerate}
    \item Generation of OCs with:
    \begin{itemize}
        \item A specified cluster formation rate (CFR).
        \item Heights following a Laplace distribution.
        \item Masses drawn from an initial mass distribution (initial cluster mass function - ICMF).
    \end{itemize}
    \item Integration of OC orbits under the MW potential.
    \item Disruption of OCs by different mechanisms: 
    \begin{itemize}
        \item Encounters with GMCs.
        \item Mass loss due to stellar evolution.
        \item Mass loss due to dynamical evolution.
    \end{itemize}
\end{enumerate}

\begin{figure*}
\centering
    \resizebox{\hsize}{!}{\includegraphics{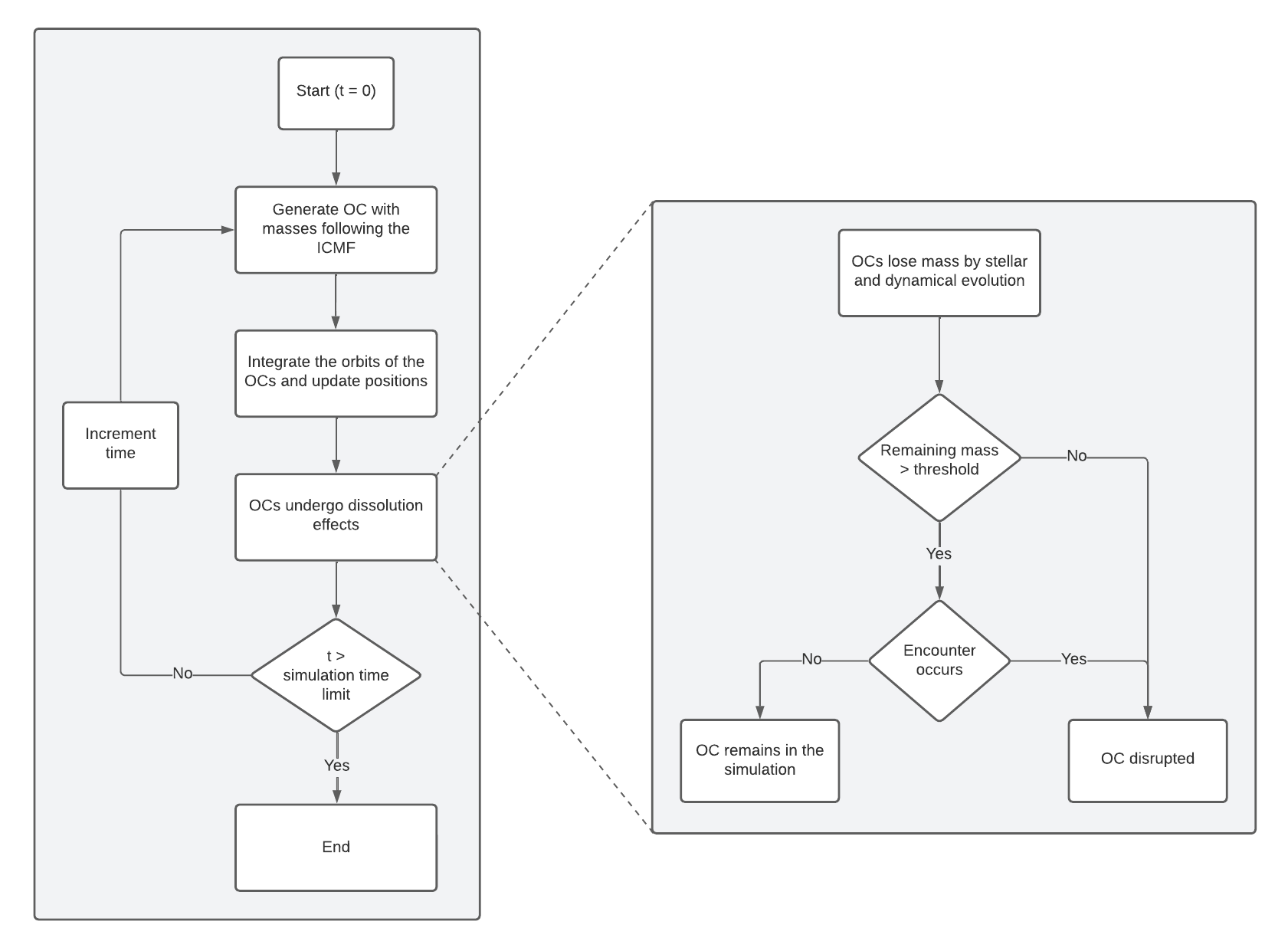}}
    \caption{Flow chart of the proposed model to explain the SH evolution. {Left: Global architecture. Right: Flow for OC dissolution processes. Includes continuous mass loss due to stellar and dynamical evolution, and disruption due to interaction with GMCs.}}
    \label{fluxograma}
\end{figure*}

The probability of the encounters depends on the vertical distance of the OCs to the GP. For simplicity, our model considers that every encounter with GMCs leads to the complete disruption of the OC. On the other hand, stellar and dynamical evolution will gradually decrease the mass of the OCs. When a minimum mass reached, the OC is considered disrupted. Figure \ref{fluxograma} shows a schematic overview of the model. {At the end of the simulation, we analysed all the surviving objects. }

In this section, we discuss the details of the model implementation. At the end of the section, we present a summarised overview of the model parameters. The inference of the parameters is presented in the next section.

\subsection{Generation of open clusters}

We consider the CFR to be constant {during the simulated period \cite{2015A&A...578A..87S, 2023A&A...669A..10Z}}. In every increment of time, the model generates 10 OCs with different masses and orbits. Each increment of time corresponds to 0.2 Myr. We note that the number of generated clusters is arbitrary and that the resulting numbers of clusters will later be scaled to match the size of the observed sample.

\subsubsection{Spatial distribution}

As was discussed in Sect. \ref{sec:ObservationalData}, the sample completeness decreases with heliocentric distance. Fig. \ref{surfacedensity} shows that the age groups younger than 1 Gyr exhibit similar declines in surface density at all probed distances. However, we note that the age group $500 < $ age (Myr) $< 1000$ {displays a somewhat more} pronounced decline at closer distances  (up $\sim$ 1  kpc) compared to the other age groups younger than $1000$ Myr. This is likely a fluctuation rising from the less populated inner distances, and thus does not differ much from the average spatial distribution of the three young age groups. Nevertheless, we have opted for a conservative approach, choosing to generate OCs following the spatial distribution of OCs with ages $< 500$ Myr.

\begin{figure*}
    \centering
    \includegraphics[width=0.49\textwidth]{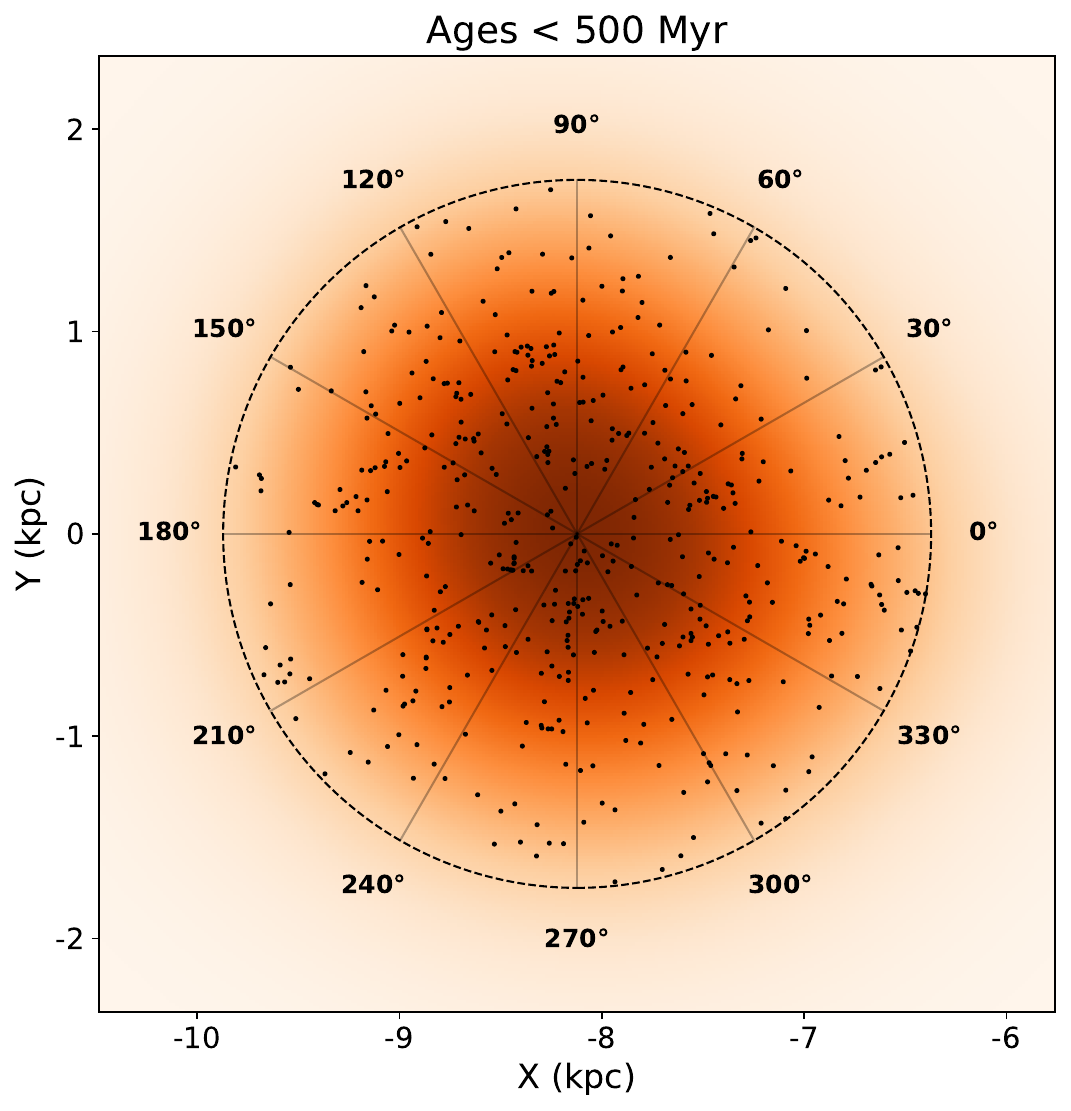}
    \includegraphics[width=0.49\textwidth]{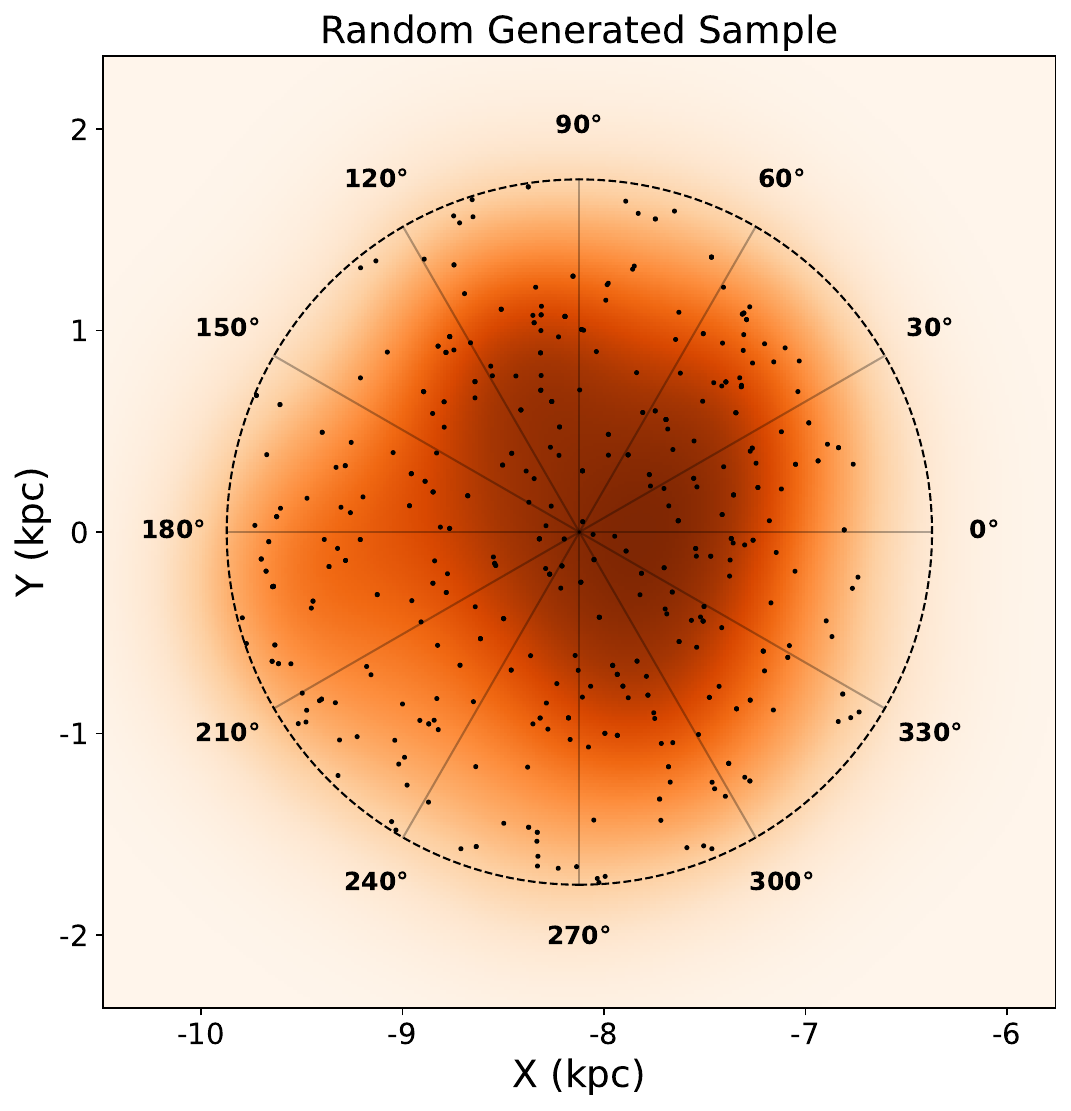}
        \caption{Observed spatial distribution of OCs and comparison with a randomly generated sample. Left: Spatial distribution of OCs projected onto the GP, for OCs ages $< 500$ Myr and $R_{\odot} < 1.75$ kpc. Right: Example of a random generated spatial distribution of OCs. Both plots include KDEs for a better comparison of the decrease in the number density with heliocentric distance.}
    \label{KDE_500_GP}
\end{figure*}

{To generate simulated OCs with appropriate spatial distributions, we began by estimating the probability distribution function (PDF) from the observational samples using a KDE. The simulated samples were then drawn from this estimated PDF. An Epanechnikov kernel with a large bandwidth of 400 pc was selected to ensure a decline in the number of OCs with heliocentric distance that was consistent with the observed sample.}
Figure \ref{KDE_500_GP} shows a comparison between the projected spatial distributions of observational data and an example of a randomly generated distribution. 

Similarly to the age groups discussed in this paper, the height distribution of OCs at birth was {drawn from} a Laplace distribution, 
as is given in Eq. \ref{eq:SH}. We designate the height distribution of the OCs at birth as the ‘initial cluster height function’ and the corresponding scale height as the ‘birth scale height’, $B_{SH}$.

\subsubsection{Initial velocities}

To integrate the OC orbits, we employed the \emph{galpy} Python package for galactic dynamics \citep{Bovy_2015}, as is further detailed in Sect.~\ref{sec:orbits}. The spatial co-ordinates were provided as $(x, y, z)$, using the galactocentric co-ordinate frame from \textit{astropy SkyCoord}. Consequently, orbit integration with \emph{galpy} requires initial values for all three velocity components. 

Although this work is concerned with vertical motions perpendicular to the GP, OCs also revolve around the centre of the MW.  Because the orbital velocity depends on the distance from the Galactic centre, it is necessary for this to be correctly attributed for each OC. \emph{galpy} allows one to calculate the $V_T$  for different galactocentric distances. Thus, the initial $V_x$ and $V_y$ for each OC can easily be derived from the initial $V_T$. In this work, we do not not consider velocity dispersions with respect to the rotation curve. 

Moreover, it is important to note that OCs form with a certain V$_\text{z}$. In this study, we consider the OCs to be 
‘dropped’ in the galaxy potential; that is, formed with V$_\text{z} = 0$. The implications of this assumption will be addressed later. In a future study, we aim to analyse the velocity profiles of young OCs, determine the distribution of their initial velocities, and compare them to observations.

\subsubsection{Initial cluster mass function}

In our model, the OCs are completely disrupted in every encounter, and thus there is no {mass dependence} in that regard. However, we also account for additional mass loss by dynamical and stellar evolution, which are mass-dependent processes \citep{Lamers2005_B}. Implementing an ICMF not only makes the model more realistic, but it also provides the means to explore in the future the impact of OCs not being completely disrupted in single encounters with GMCs.

Following previous works \citep{Lada_2003,2003A&A...397..473B}, we adopted an ICMF described by a power law: $dN/dM \propto M^{-\alpha} \text{ with } \alpha \approx 2$. To generate OCs following this mass distribution, we used the inverse transform sampling method. The masses were obtained by substituting a randomly generated number between 0 and 1 in the following expression:
\begin{equation}
    M = \bigg(\frac{1}{M_{min}} - \mu\text{ }\bigg(\frac{1}{M_{min}} - \frac{1}{M_{max}}\bigg)\bigg)^{-1}
.\end{equation}
We adopted $M_{min} = 10^2M_\odot$ and $M_{max} = 3\cdot10^4M_\odot$ \citep[as suggested by ][for clusters in the solar neighbourhood]{LamersGieles2006}.

\subsection{Orbit integration\label{sec:orbits}}

As has already been referred to, OC orbits were integrated using \emph{galpy} \citep{Bovy_2015}. We selected the axisymmetric MW potential, MWPotential2014, recommended by \citet{Bovy_2015}. It  has three components: a bulge with a density profile given by a power law with an exponential cut-off, a disc described by the Myamoto-Nagai potential \citep{MiyamotoNagai}, and a dark matter halo modelled with a Navarro-Frenk-White potential (NFW; \citep{NFW}). We used the default values of \emph{galpy} for each component \citep[see][for detailed description of the adjustment of the different components.]{Bovy_2015}.

By specifying the initial conditions, one can integrate the orbits for any evolution time with the desired time steps. The orbits can be initialised using different co-ordinate frames, containing positional co-ordinates and velocity components. As {previously}, we used the galactocentric frame from \emph{astropy SkyCoord}. It is worth noting that the default values for the position of the Sun, which are used to scale the potential, differ between \emph{galpy} and \emph{astropy}. To ensure consistency, as is outlined in \emph{galpy} documentation, one should specify in the \emph{astropy} co-ordinate frame the following: the Sun's distance to the Galactic centre, $R_{\odot} = 8$ kpc; the Sun's height, $Z_{\odot} = 15$ pc; and velocity components, $(V_x, V_y, V_z) =$ (10, 235, 7) km/s.

\subsection{Open cluster disruption mechanisms}

Here, we describe how encounters with GMCs combined with stellar and dynamical evolution were included in the model. We started by considering an OC to be disrupted if any of the following conditions were met: 
\begin{itemize}
    \item the OC encounters a GMC.
    \item the OC reaches a minimum mass threshold (due to the gradual mass loss caused by stellar and dynamical evolution).
\end{itemize}  

The minimum mass threshold that we defined depends on the observational capacity of detecting an OC. Clusters with low masses, and thus a very small number of stars, are very hard to detect due to their low luminosities and density contrast with the field along the line of sight. In \citet{PiskunovTidalMasses}, the authors estimate the masses of 236 OCs using the tidal radius. The minimum mass estimated was $\sim 18 M_{\odot}$. Recently, \citet{Almeida_2023} using \textit{Gaia} early DR3 determined masses of 773 OCs within $\sim 4$ kpc from the Sun. {The minimum mass in their sample is} $\sim 60 M_{\odot}$. Also, our own work \citep{Duarte} indicates the presence of a few stellar aggregates with masses down to  $\sim 20 M_{\odot}$ although almost all OCs have masses in excess of $\sim 67 M_{\odot}$. Here, we adopted the conservative value of $\sim 18 M_{\odot}$ for our minimum mass threshold. In any case, given the quick dispersal times of low-mass clusters \citep{LamersGieles2006}, adopting any of these values does not make a difference, {as we have confirmed by running the model with minimum mass thresholds between $18-60 M_{\odot}$.}

\subsubsection{Encounters with giant molecular clouds\label{sec:model_GMC}}
The implementation of this mechanism in our model has the following assumptions:
\begin{enumerate}
    \item The vertical distribution of GMCs can be approximated well by a Laplace distribution.
    \item A homogeneous distribution of GMCs around the GP.
    \item Every encounter leads to the complete disruption of the OC.
\end{enumerate}
Although GMCs exhibit a clumpy distribution around the GP at smaller scales of $\sim 100$ pc \citep[e.g.][]{2001ApJ...547..792D}, our model makes the simplifying approximation of an average homogeneous distribution. This should not influence the results on the kiloparsec scales addressed in this study. 

Considering the previous assumptions, the probability of an encounter is associated with the spatial distribution of the GMCs: 
\begin{equation}
    \label{p_function}
    p(\text{z};\text{D}_{\text{SH}}) = p_0 \hspace{0.06cm}\text{exp}\left({-\frac{\vert Z \vert}{\text{D}_{\text{SH}}}}\right)
.\end{equation}
Here, p$_0$ is interpreted as the scale factor that regulates the intensity of the disruption of OCs due to encounters with GMCs. The $D_{SH}$ stands for dissolution scale height and represents the effective SH of the disruption due to encounters. Higher values of $D_{SH}$ imply that OCs at greater distances to the GP are more affected by encounters with GMCs. 

At each time step, the model assesses the probability of an encounter for each OC. Subsequently, a uniformly distributed random number between 0 and 1 is drawn and, if this number is higher than the probability of an encounter, the cluster is considered disrupted.

\subsubsection{Stellar and dynamical evolution\label{subsect:sde}}

{In contrast to encounters with GMCs, stellar and dynamical evolution lead to a gradual loss of mass over time. To implement these effects, we employed the expression for cluster mass loss provided by \citet{Lamers2005_B}, which combines contributions from both stellar and dynamical evolution (their Eq. 6):
}

{
\begin{equation}
    \mu(t; M_i) \equiv \frac{M(t)}{M_i} \approx \left [ (\mu_{ev}(t))^{\gamma} - \frac{\gamma t}{t_0} \left( \frac{M_{\odot}}{M_i} \right)^{\gamma} \right]^{1/\gamma}
    \label{tidal_sv}
.\end{equation}
}

{Here, $\mu(t; M_i)$ represents the mass loss rate at time $t$ for an OC with initial mass $M_i$. The first term describes mass loss due to stellar evolution, where $\mu_{ev}(t)$ denotes the fraction of the initial OC mass remaining at a given age, $t$, assuming that mass loss occurs solely due to stellar evolution. The second term accounts for mass loss from dynamical evolution, characterised by parameters $t_0$ and $\gamma$. For a detailed description, we refer readers to \citet{Lamers2005_B}. For simplicity, the meaning of these parameters can be understood through the expression for the disruption time of a star cluster \citep{BaumgardtMakino, BL03, Lamers2005_B}:}

{
\begin{equation}
    t_{dis} = t_0 \left( \frac{M_i}{M_{\odot}} \right)^\gamma
\end{equation}
}

{Here, $t_{dis}$ is the disruption time for a cluster with initial mass $M_i$, and $\gamma$ indicates the mass dependence of the disruption time. N-body simulations \citep{BaumgardtMakino} and observations in the solar neighbourhood and other galaxies \citep{Lamers2005_A} consistently suggest a value of $\gamma \approx 0.6$. In this work, we adopt $\gamma = 0.62$, which was determined by \citet{Lamers2005_B}. }

{The parameter $t_0$ sets the disruption timescale and depends on the tidal field experienced by the cluster as well as the ambient density at its position within a galaxy. Although previous studies have estimated $t_0$, we retained it as a free parameter in our model for the following reasons. \citet{Lamers2005_B} estimated $t_0$ from observations, assuming that the total observed mass loss in OCs is described by Eq.~\ref{tidal_sv}. In contrast, our model explicitly separates the contributions of stellar and dynamical evolution -- modelled by Eq.~\ref{tidal_sv} -- from interactions with GMCs, which are treated in the manner described in the previous section. Consequently, our estimates of $t_0$ should be larger, since they exclude the additional destruction due to GMC interactions. Furthermore, the observational sample used by \citet{Lamers2005_B} was based on the OC catalogue of \citet{KarchenkoCatalogue}. Since then, \textit{Gaia} has significantly improved the census of OCs. By using the \textit{Gaia}-based catalogue of \citet{DiasCat}, we expect to achieve more accurate determinations.}

\section{Parameter inference}

{In this section, we discuss the inference of our model's parameters and the methods used to make a comparison with observations and make these inferences. We begin by summarising the free parameters of the model along with their descriptions:}
\begin{itemize}
    \item B$_{\text{SH}}$ (in parsecs) - Scale height of OCs at birth.
    \item $p_{0}$ (per million years) - Scale factor of the probability of encounters with GMCs.
    \item $D_{\text{SH}}$ (in parsecs) -  Scale height of the disruptive effects of encounters with GMCs.
    \item $t_{0}$ (in years) - Timescale of the disruption due to dynamical evolution.
\end{itemize}

{
An initial exploratory analysis revealed that simultaneous optimisation of these parameters leads to some degeneracy in the solutions, with variations in $B_{\text{SH}}$ being compensated for by adjustments in $p_{0}$ and $D_{\text{SH}}$. This behaviour is expected, since an increase in a birth parameter can be partially offset by changes in the disruption parameters. To mitigate this issue, we separately inferred the birth and disruption parameters. The next section focusses on the birth scale height, $B_{\text{SH}}$, while the subsequent section addresses the remaining disruption parameters: $p_{0}$, $D_{\text{SH}}$, and $t_{0}$.}

\subsection{Birth scale height}

The first parameter to consider is the birth scale height, $B_\text{SH}$. 
Ideally, it would be determined from observations of spatial distribution of clusters at birth time. However, this is hampered by several difficulties. The first one is the small numbers of newly born clusters (e.g. ones younger than 1 Myr) in the solar neighbourhood, which would provide poor statistics. Furthermore, most very young clusters are still deeply embedded in their parent molecular clouds \citep[e.g.][]{Lada_2003} and only observable at infrared wavelengths; in other words, out of \textit{Gaia}'s reach. Given this situation, we need to consider a somewhat older sample and take into account the evolutionary effects that might affect the inference of the  $B_\text{SH}$, as we do below.

\begin{figure}
\centering
    \resizebox{\hsize}{!}{\includegraphics{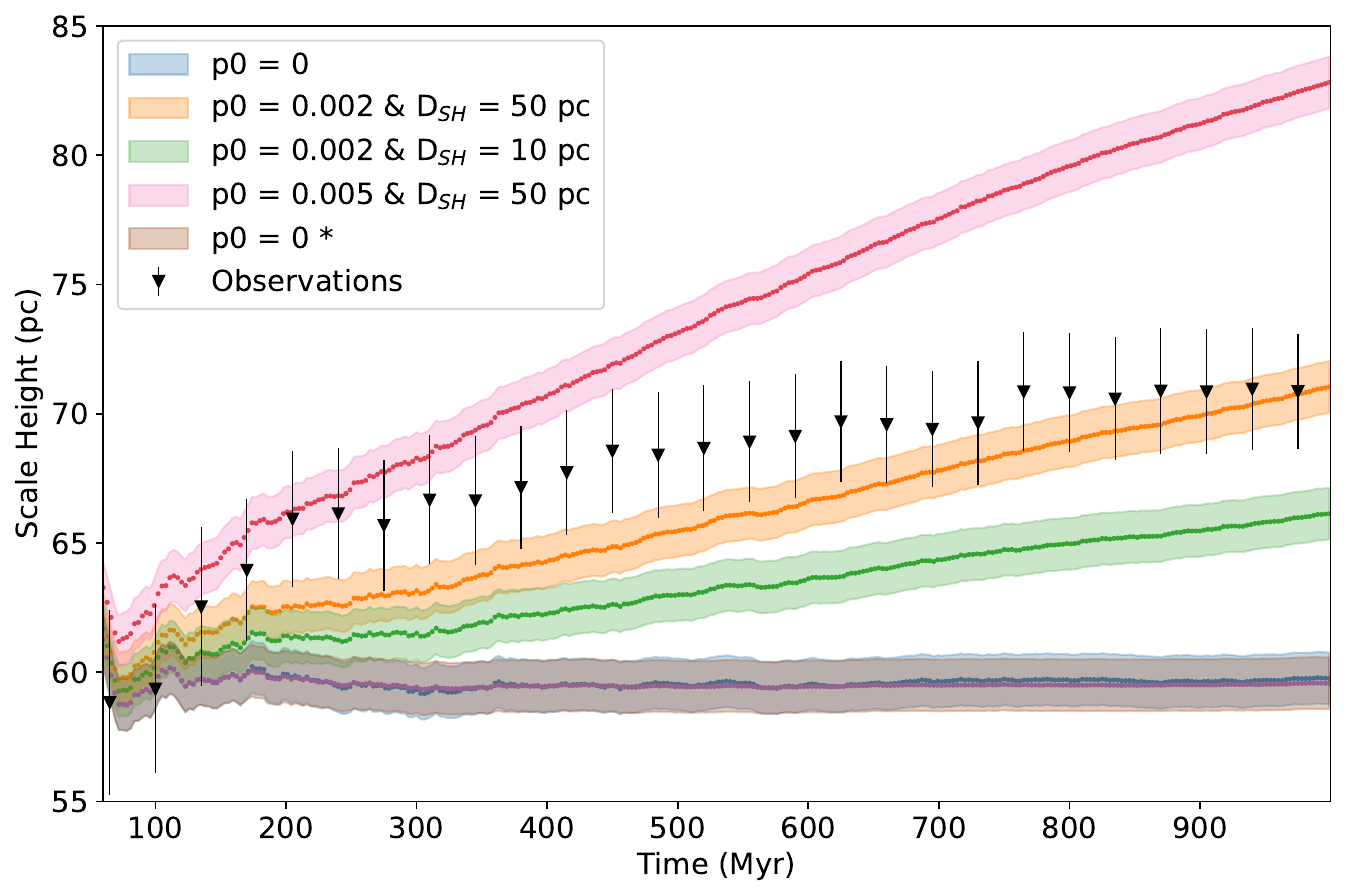}}
    \caption{Comparison of the SH evolution for different combination of parameters with $B_\text{SH} = 92$ pc and for the observations. The black triangles are the mean SH obtained from 1000 bootstrap runs on the observational sample and the error bars are the standard deviations; the blue curve corresponds to a run without any disruption mechanism activated; the brown curve includes only mass loss due to stellar and dynamical evolution; and the pink, orange, and green curves include only disruption due to encounters with GMCs, with different intensities and dissolution scale heights. The curves correspond to the averages of ten independent runs and the shaded regions are the standard deviations.}
    \label{BSH}
\end{figure}

To better understand what might be a good age limit to make this comparison, Fig. \ref{BSH} shows for a $B_\text{SH} = 92$ pc, the cumulative SH evolution of four runs of the model with different parameter combinations, along with the observations. This gives a first glimpse of the effects of each parameter on the OC scale height evolution. It is immediately seen that the SH grows with age for the data as well as for the simulations (except for the simulation with no disruption from GMCs, in brown and blue). Although there are small fluctuations in the modelled SH evolutions for younger ages, the SH grows systematically after 150 Myr. 

For times $< 100$ Myr, the curves overlap almost perfectly, except for the pink curve, which is slightly above. This indicates that the dissolution mechanisms considered here do not have a significant effect in that time frame. Thus, we adjusted the $B_{SH}$ to match the SH of the observations considering only the OCs younger than 100 Myr. Considering the 1.75 kpc cylindrical cut, the catalogue has 228 OCs with an age $< 100$ Myr and a corresponding SH of $59 \pm 3$ pc. This value was calculated by using 1000 bootstrap runs. We note, however, that the  $B_{SH}$ will necessarily be larger than the observed SH. There are two reasons for this:

\begin{enumerate}
    \item As the OCs form, the gravitational potential makes them fall towards the GP. Therefore, when we observe the population of OCs, on average, the heights will be smaller than the initial heights at which the OCs were dropped.
    
    \item Initial $V_z \neq 0$ reduces this effect as the additional kinetic energy increases the vertical amplitudes of the orbits, allowing them to reach larger heights. This allows for the possibility of the OCs being observed at larger heights than the ones at which they were formed. Because in our model the OCs are formed with the initial $V_z = 0$, at any given time, the heights of the OCs will necessarily be equal to or smaller than their initial heights.
\end{enumerate}
Using the model with no disruption activated, we find that the relation between the observed SH at 100 Myr ($O_{SH}$) and the $B_{SH}$ is linear: $O_{SH} = m B_{SH}$, with $m = 0.639 \pm 0.001$. As was mentioned above, the SH for OCs with age $< 100$ Myr is $59 \pm 3$ pc. This led to us adopting a value of $B_{SH}= 92 \pm 5$ pc, where the uncertainty was calculated using error propagation.

We remark that the actual $B_{SH}$ of OCs may be smaller than this value, due to initial velocities as was discussed, but also because we might be missing some very young OCs as their light gets blocked by the gas of the clouds in which they formed. This means that we shall miss some low-altitude clusters, artificially increasing the $O_{SH}$.

\subsection{Disruption parameters}

\begin{figure}
\centering
    \resizebox{\hsize}{!}{\includegraphics{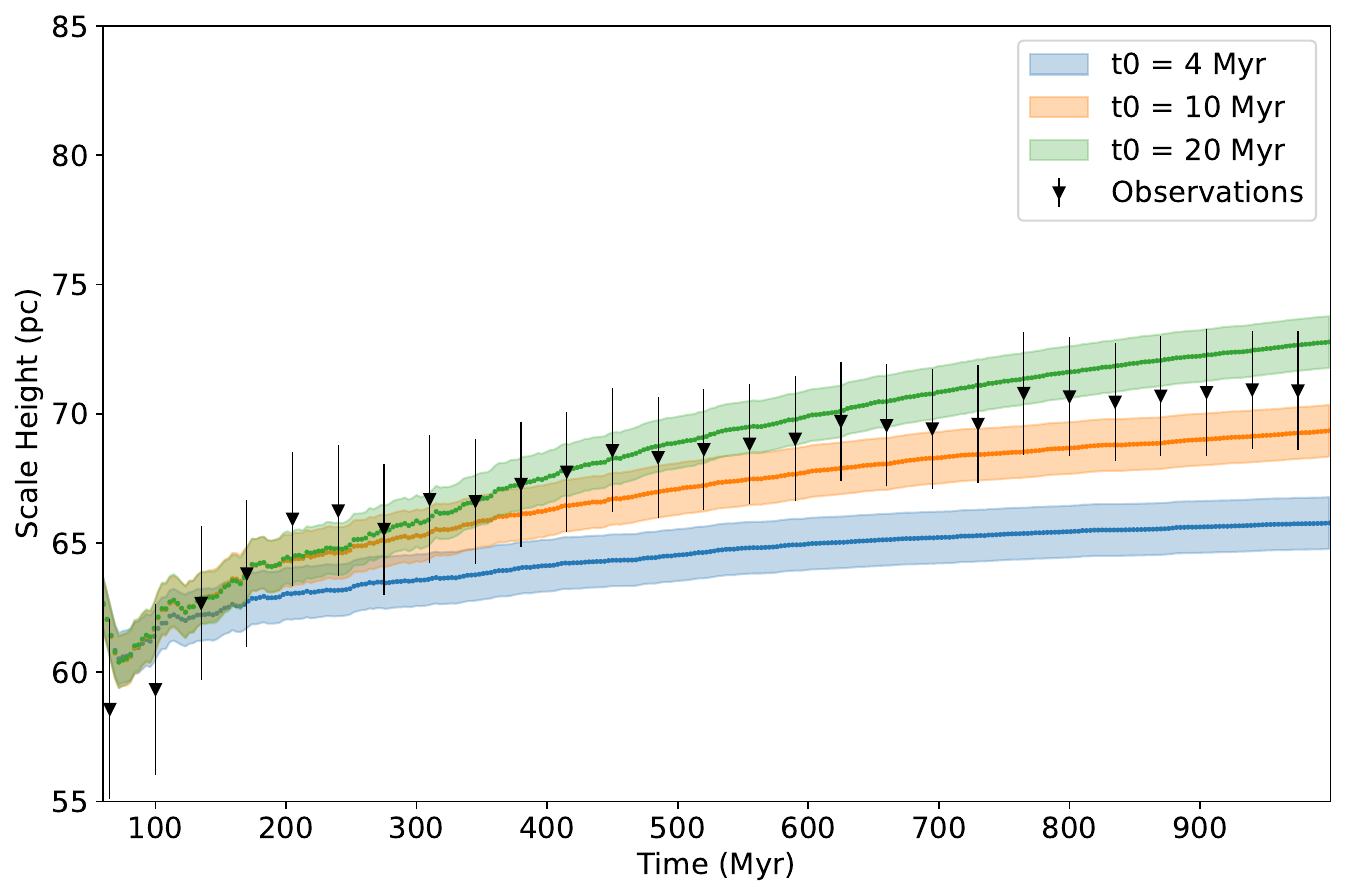}}
    \caption{Comparison of the SH evolution for runs with different values of $t_0$ ($B_\text{SH} = 92$ pc, $p_0$ = 0.004, and $D_{SH}$ = 30 pc). The curves correspond to the averages of ten independent runs and the shaded regions are the standard deviations. The black triangles are the mean SH obtained from 1000 bootstrap runs on the observational sample and the error bars are the standard deviations.}
    \label{degeneracy}
\end{figure}

With the $B_{SH}$ fixed, we can now explore how the other parameters affect the SH evolution. We start by noting that although Fig. \ref{BSH} shows that $t_0$ does not affect the SH evolution of the simulations without GMC encounters, the situation changes when interactions with GMCs are considered. Despite $p_0$ and $D_{SH}$ being the parameters directly related to the encounters with GMCs, the study of these parameters cannot be made independently of $t_0$. This is easily understood with an extreme example: a very strong mass loss by stellar and dynamical evolution  will disrupt OCs before they have the opportunity to encounter GMCs. Thus, if the OCs lose more mass due to these mechanisms (decrease in $t_0$), the probability of encounters has to increase, either by increasing $p_0$ and/or $D_{SH}$, allowing the OCs to encounter GMCs before being disrupted. The same logic applies to the opposite scenario of the OCs {losing} less mass due to stellar and dynamical evolution (increase in $t_0$). For clarity, Fig. \ref{degeneracy} shows how different values of $t_0$ affect the SH evolution. The different curves have the same $p_0$ and $D_{SH}$.

Additionally, even though $p_0$ and $D_{SH}$ are conceptually different, different combinations of these parameters lead to similar rates of disruption by encounters with GMCs. Thus, the model displays some degeneracy in its parameters. To address this, we compared not only the SH evolution but also the total number of clusters in existence at different ages. We conducted a grid search to find the best solutions, assessing their fit quality in the following way: 

\begin{enumerate}
    \item We compared the total number of OCs that are younger than a given age for the ages between 10 and 1000 Myr with a 20 Myr step. The decision to limit the comparison to ages up to 1000 Myr was motivated by the distinct trends observed in figures \ref{galacticPlane} and \ref{surfacedensity}, in which the numbers of OCs exhibit a distinct decrease for ages greater than 1000 Myr. Consequently, OCs were generated based on the spatial distribution of the younger population (ages $< 500$ Myr) to accurately reflect the evolution of numbers, leading to the exclusion of ages older than 1000 Myr from this analysis. For a proper comparison, the number of OCs from the results needs to be scaled to the observations. To estimate the scaling factor, we applied a least squares estimation. The optimum of the fit was calculated using the log-likelihood estimation. 
    Because we are comparing cumulative distributions, for which the number of OCs in each subsequent bin is dependent on the number of the previous bin, the covariance matrix needs to be included into the maximum likelihood estimation. The log-likelihood, for each independent run, is given by \citep{james_introduction_2013}: 
    
    \begin{equation}
       l(x; \mu, \Sigma, N) = -\frac{\text{N}}{2} \ln{(2\pi)} - \frac{\text{N}}{2} \ln \vert \Sigma \vert - \frac{1}{2} (x - \mu)^{\text{T}} \Sigma^{-1} (x - \mu)
    ,\end{equation}
    where N is the total number of bins, $\Sigma$ is the covariance matrix, $\mu$ is a vector with the expected values -- the observed values for each bin -- and $x$ is the vector with the number of OCs in each bin from the simulation. 


    
    \item We compared the SH evolution from the simulations and the observations in the proposed age groups. Furthermore, we included additional age intervals to make a more robust estimation (the additional age intervals are shown in Sect. ~\ref{sec:SH_evol}). For the SH, we still considered OCs with ages up to 2000 Myr. Their vertical distribution should be independent of the total number of OCs. Once again, we assessed the quality of the fit by calculating the log-likelihood estimation. The covariance matrix was no longer required for the SH, since we did not use a cumulative distribution. Thus, the log-likelihood can be simplified:

    \begin{equation}
        l(x; \mu, \sigma, N) = -\frac{\text{N}}{2} \ln{(2\pi)} - \sum_{i=1}^{N} \left [ \frac{\text{1}}{2}  \ln{\vert \sigma_i \vert} +  \frac{(x_i - \mu_i)^2}{2\sigma^2} \right]
    .\end{equation}

    Here, $\mu_i$ is the SH from the observations for each bin and $\sigma_i$ is the corresponding standard deviation. Both values were obtained from 1000 bootstrap runs. $x_i$ represents the average value of the SH, for ten independent runs, in the corresponding age bin and for a specific combination of parameters.
    
    \item The final score for the goodness-of-fit is given by the sum of both log-likelihoods: the age and numbers evolution and the SH evolution. Because the two log-likelihoods have different ranges of values, we scaled each one between 0 and 1.
\end{enumerate}

\begin{figure}
\centering
\resizebox{0.955\hsize}{!}{\includegraphics{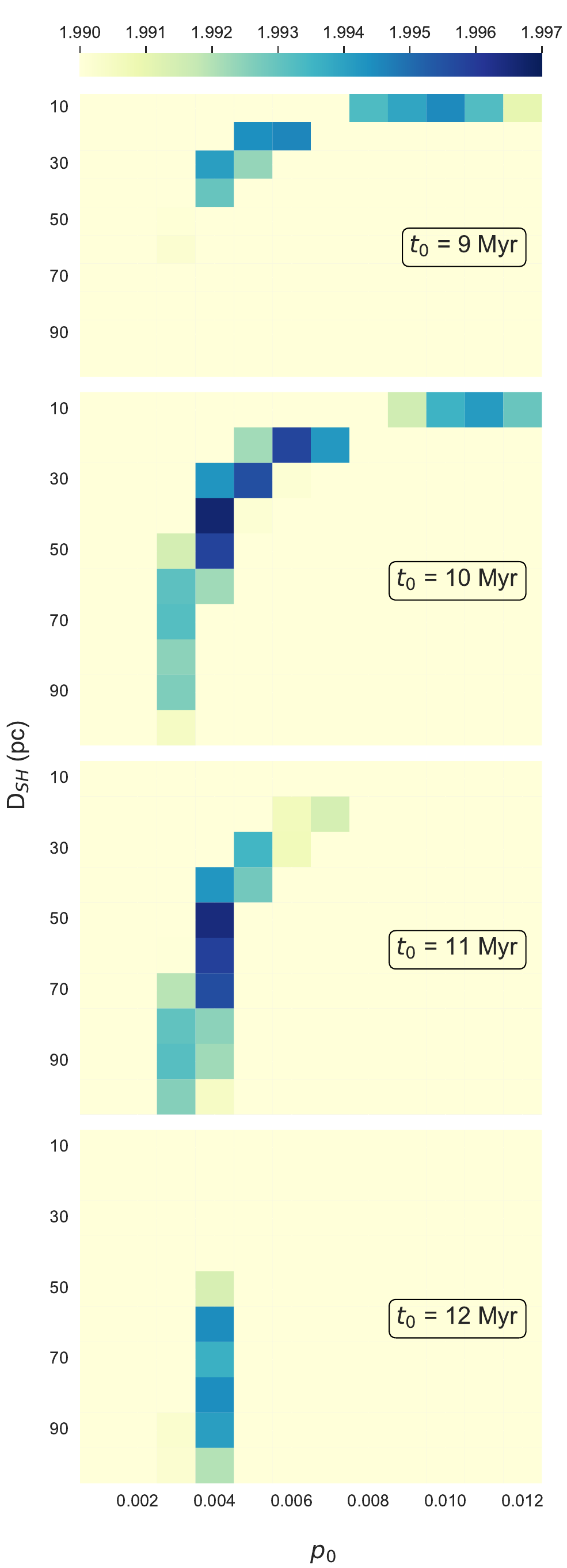}}
    \caption{Heat maps based on the combined log-likelihood estimations for different combinations of parameters.}
    \label{fig:heatmaps_death}
\end{figure}

Figure~\ref{fig:heatmaps_death} presents heat maps for multiple combinations of $p_0$ and $D_{SH}$. Each heat map corresponds to a specific value of $t_0$. Darker colours indicate a better fit to the observations. It is important to mention that the implementation of the covariance matrix in the maximum likelihood estimation was fundamental to remove, partially, the degeneracy in the parameters space. We note the usual practice in the literature is to fit cumulative distributions without considering that cumulative bins are correlated.  

\subsubsection{{$t_0$ and $D_{SH}$}}
{As is seen in Fig.~\ref{fig:heatmaps_death}, the best fits are achieved for $t_0 = 10-11$ Myr. 
For $t_0 = 10$ Myr, we find $D_{SH} \sim 30-50$ pc, with the optimal value at 40 pc.
For $t_0 = 11$ Myr, we find $D_{SH} \sim 40-70$ pc, with the optimal value at 50 pc.
We argue that $t_0 = 11$ Myr is the best solution.}
{In this work, we have considered the OCs to have an initial $V_z = 0$ (i.e. they are dropped at birth). This tends to lead to underestimated values of $D_{SH}$. To understand why, we can consider an initial $V_z \neq 0$. The initial velocity will increase the amplitudes of the orbits, meaning that they will arrive at the GP with higher speeds, and thus spend less time at closer distances from the GP, leading to less encounters. Therefore, for the same number of encounters, initial velocities will require a thicker $D_{SH}$ in order to be equally exposed to the same overall destructive effect.}

{We compared our result for $t_0$ with the ones obtained from N-body simulations. The extensive grid of simulations conducted by \citet{BaumgardtMakino} suggests a dissolution timescale of $t_0 = 18.1$ Myr. In their work, the value is given as $t_4 = 6.9$ Gyr, which we converted to $t_0$ using the conversion formula, $t_4^\text{total} = 6.60 \times 10^2 t_0^{0.967}$, provided by \citet{Lamers2005_B}. However, \citet{LamersGieles2006} noted that the results from \citet{BaumgardtMakino} overestimate the dissolution timescale for OCs with masses below 4500 $M_{\odot}$ up to a factor of approximately two, which applies to nearly all of the objects in our sample. This makes our result consistent with those from the N-body simulations.}

{Regarding $D_{SH}$, although the SH of the distribution of GMC central positions is known to be small, roughly 20-40 pc \citep{Stark&Lee2005}, their sizes can range from a few parsecs to $\sim$ 200 pc \citep{Lada&Charles2020}. While the effective SH considering the dimensions of the clouds is not catalogued, the half-thickness at half-intensity of the CO layer is found to be $\sim$ 70 pc in the solar vicinity \citep{1994ApJ...433..687M}. Given that the $D_{SH}$ in the model represents the effective scale height of disruption by encounters, we find that $D_{SH} \sim 40-70$ pc, corresponding to $t_0 = 11$ Myr, is in good agreement with the measurements mentioned above.}

\subsubsection{{$p_0$}\label{sec:p0}}

{As is shown in Fig.~\ref{fig:heatmaps_death}, the best fits were obtained for $p_0 = 0.004$. To compare this value with other observational constraints, we recall that $p_0$ represents the intensity of the disruption effect (see Sect.~\ref{sec:model_GMC}). Specifically, it denotes the probability of an OC encountering a GMC at $Z = 0$ (where the effect is maximal) per million years. Given a dissolution scale height ($D_{SH}$ parameter), the average time to an encounter with a GMC, of a cluster born at height $Z_B$, can be determined using the following expression:}

\begin{equation}
     \approx T \times \left[\int_0^T p_0\text{exp}\left(-\frac{\vert Z(t)\vert}{D_{SH}}\right)
    dt \right]^{-1}
,\end{equation}

{where $T$ is the vertical period of the orbit and $Z(t)$ is the time-dependent height of the object. We note that $T$ and $Z(t)$ depend on the Galactic potential, and thus the integral will in general need to be integrated numerically. For an OC in the solar neighbourhood born at a height of $Z \sim 100$ pc, we find that the average time to an encounter with a GMC is $702 \pm 23$ Myr. We chose 100 pc for this example as it approximates the current maximum height of the Sun, based on default \emph{galpy} parameters.}

{This average encounter time aligns well with the results of \citet{2019MNRAS.489.5165K}, who determined that the Sun undergoes one encounter every $625^{+2700}_{-280}$ Myr. Such a remarkable agreement with an independent result, which focusses solely on single stars and does not consider GMC-induced destruction effects, provides strong support for our model.}

\section{Discussion}
\subsection{Scale height evolution \label{sec:SH_evol}}
\begin{figure}
\centering
     \resizebox{\hsize}{!}{\includegraphics{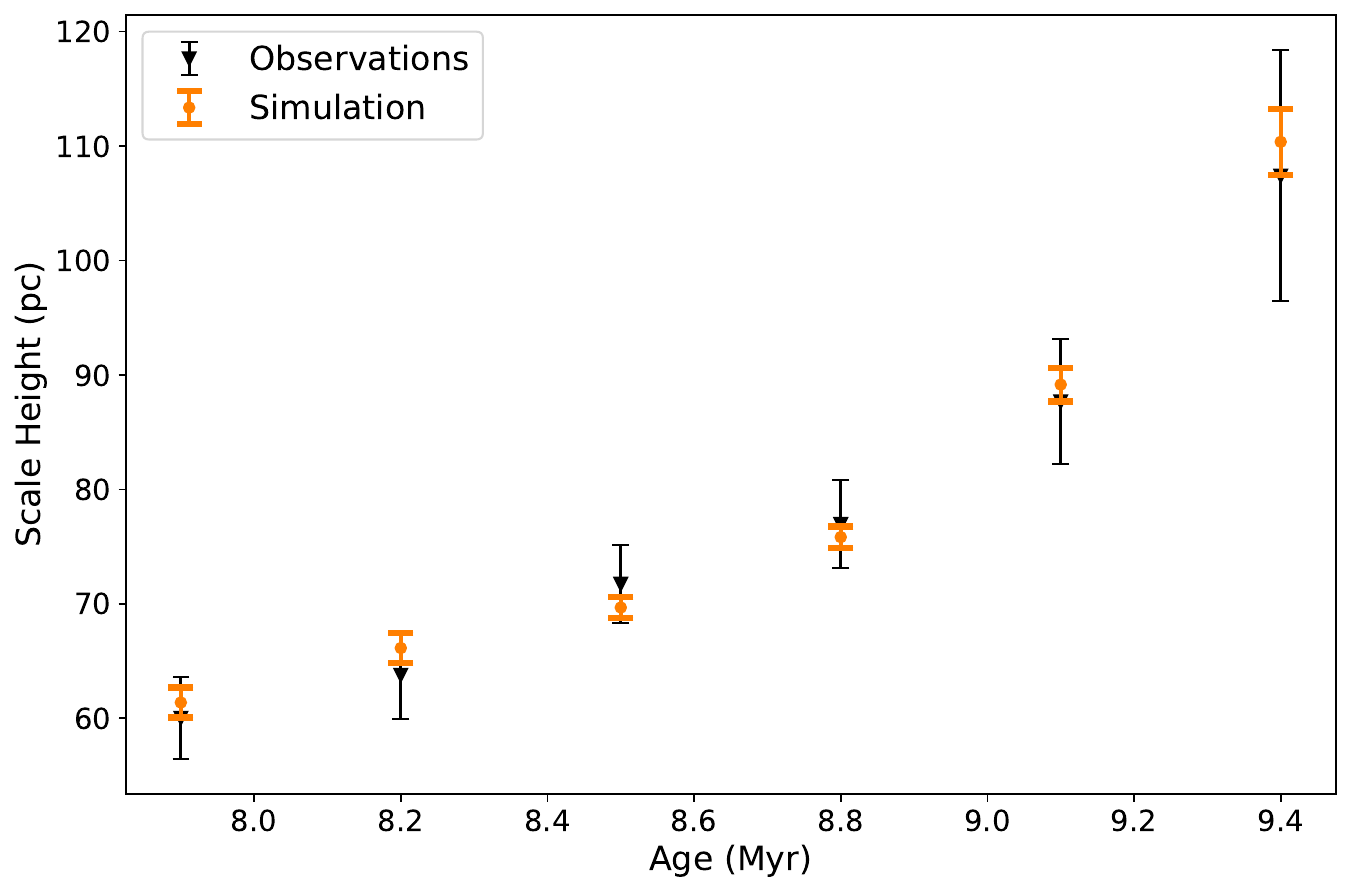}}
    \caption{Comparison of the SH evolution predicted by the model (orange dots) with the observations. The model parameters used were:  
 $B_{SH} = 92$ pc, $D_{SH} = 50$ pc, $p_0 = 0.004$, and $t_0 = 11$ Myr.}
    \label{SH_evol}
\end{figure}

Figure \ref{SH_evol} compares the SH evolution seen in the data to the average simulation results using our adopted parameters listed in Table \ref{table:combined_parameter_summary}. {The figure displays the results in overlapping bins in the range log(age) = 7.9-9.4, separated by 0.3 dex. The bins have a width of log(age) = 0.6 dex.} The error bars of the data were obtained from the standard deviations of 1000 bootstrap runs and the corresponding SH is the average of those runs. The SHs of the simulations were obtained from the average of ten independent runs and the error bars are the corresponding standard deviations of the SH for those runs. Overall, the agreement is excellent.

\begin{table}
\caption{Best combination of parameters and their standard deviations.}
\label{table:combined_parameter_summary}
\centering
\begin{tabular}{cccc}
\hline\hline
Parameter & Value & Standard Deviation \\ \hline
$B_{SH}$ (pc) & 92 & -- \\
$D_{SH}$ (pc) & $50.0$ & $7.6$  \\
$p_0$ & $0.0040$ & $0.0002$ \\
$t_0$ (Myr) & 11  & -- \\ \hline
\end{tabular}
\end{table}

\begin{figure*}
    \centering
    \resizebox{0.49\hsize}{!}{\includegraphics{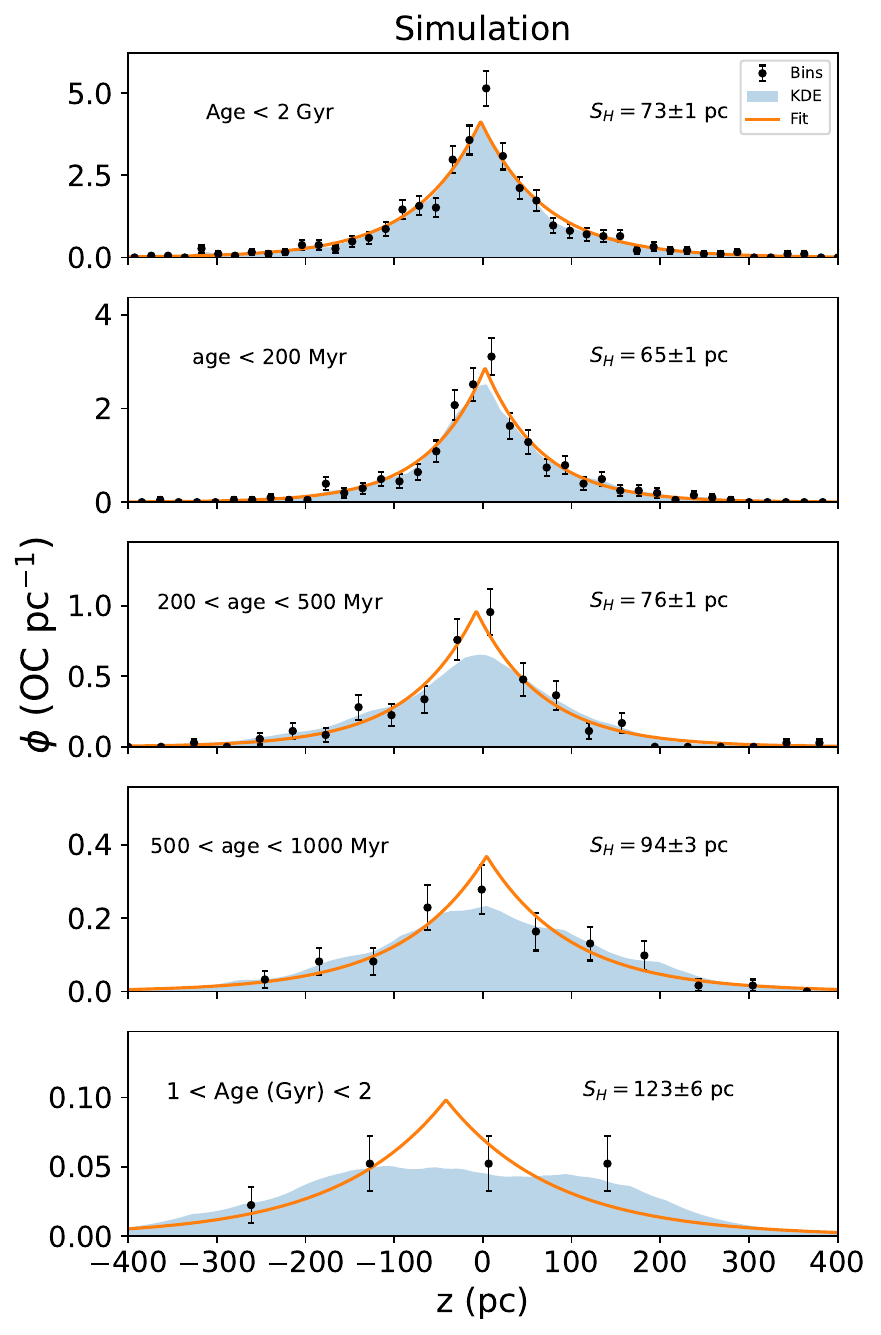}}
    \resizebox{0.49\hsize}{!}{\includegraphics{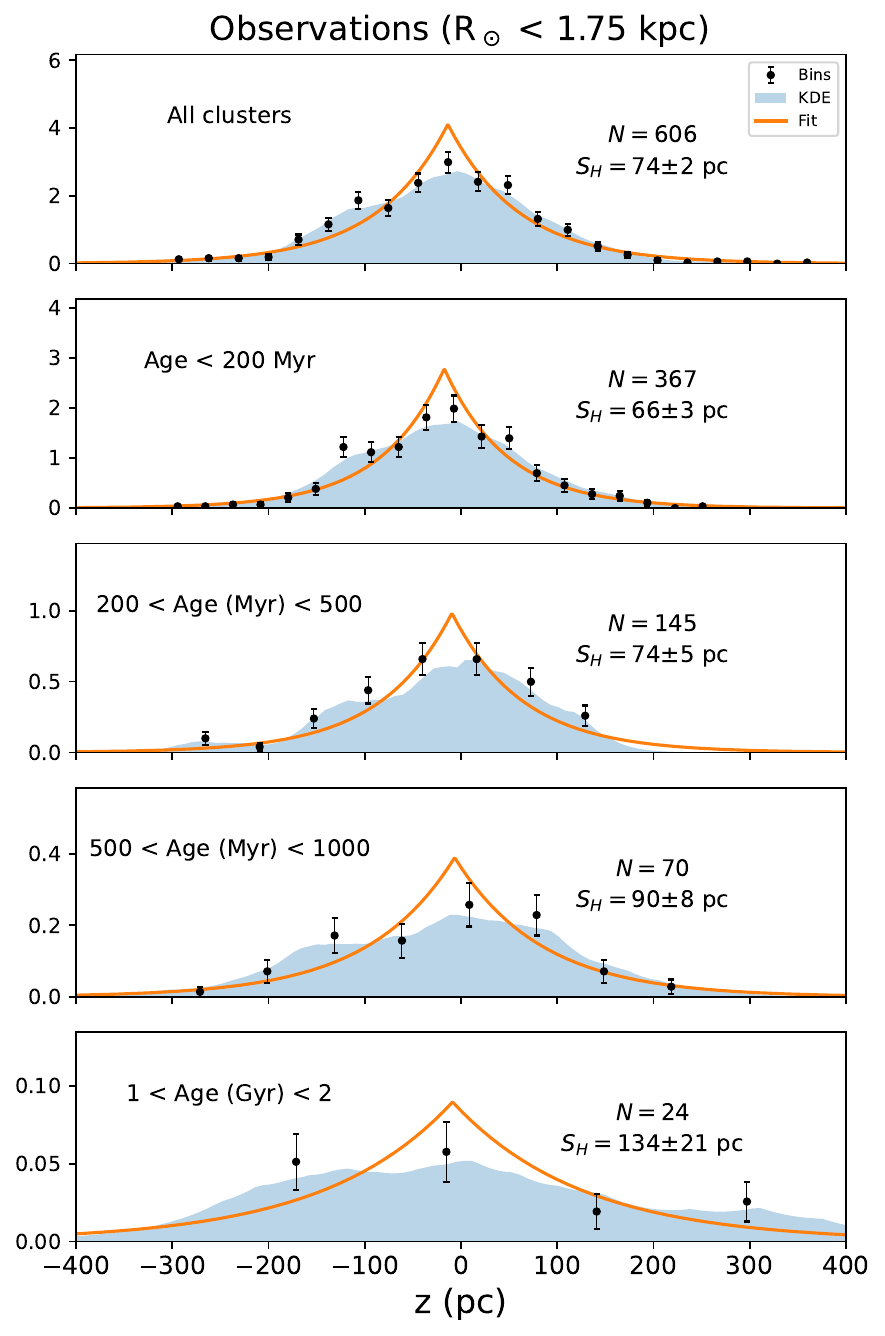}}
    \caption{Comparison of the SH in the considered age groups. Left: SH of the different age groups predicted by the model using $B_{SH} = 92$ pc, $D_{SH} = 50$ pc, $p_0 = 0.004$, and $t_0 = 11$ Myr. Right: SH of the observational data \citep{DiasCat}. The filled curve represents a KDE using an exponential kernel and the solid lines are fits with the form of Eq. \ref{eq:SH}.}
    \label{SH_KDE}
\end{figure*}

Following the traditional representation made in \citep{JanesPhelps1994} and \citet{Bonatto2006}, Fig.~\ref{SH_KDE} compares, for several age groups, the SH output from the model to observed values. The KDEs of the simulations were made by randomly selecting N elements from the resulting sample after running the model, where N corresponds to the number of OCs from the observations in the corresponding age group. The adjustment is remarkable for ages $< 1$ Gyr. For ages $> 1$ Gyr, the results are still within the uncertainties of the observations. 

This difference might be caused by missing some old OCs close to the GP. Detecting older OCs is more challenging due to their fainter nature, making them harder to observe, especially at lower heights where field confusion and light absorption by interstellar dust are more prominent. Even though the decrease in the completeness of the older age group is less accentuated  than that observed for the younger age groups (figure~\ref{surfacedensity}), this is a relatively small sample, with only 24 OCs. Thus, missing just a few old OCs close to the GP will significantly increase the determination of the observed SH. 

Aside from the observational difficulties, it is possible that the discrepancy for the older age group arises from the simplicity of the model. The assumption that disruption by encounters is independent of OC masses, and other intrinsic properties (e.g., radius, density, internal velocity dispersion), could potentially impact the results. Moreover, not considering initial vertical velocities affects the dynamics of the orbits of the OCs, which likely impacts the overall evolution of the SH.

\subsection{Age and open cluster number evolution}

\begin{figure}
\centering
    \resizebox{\hsize}{!}{\includegraphics{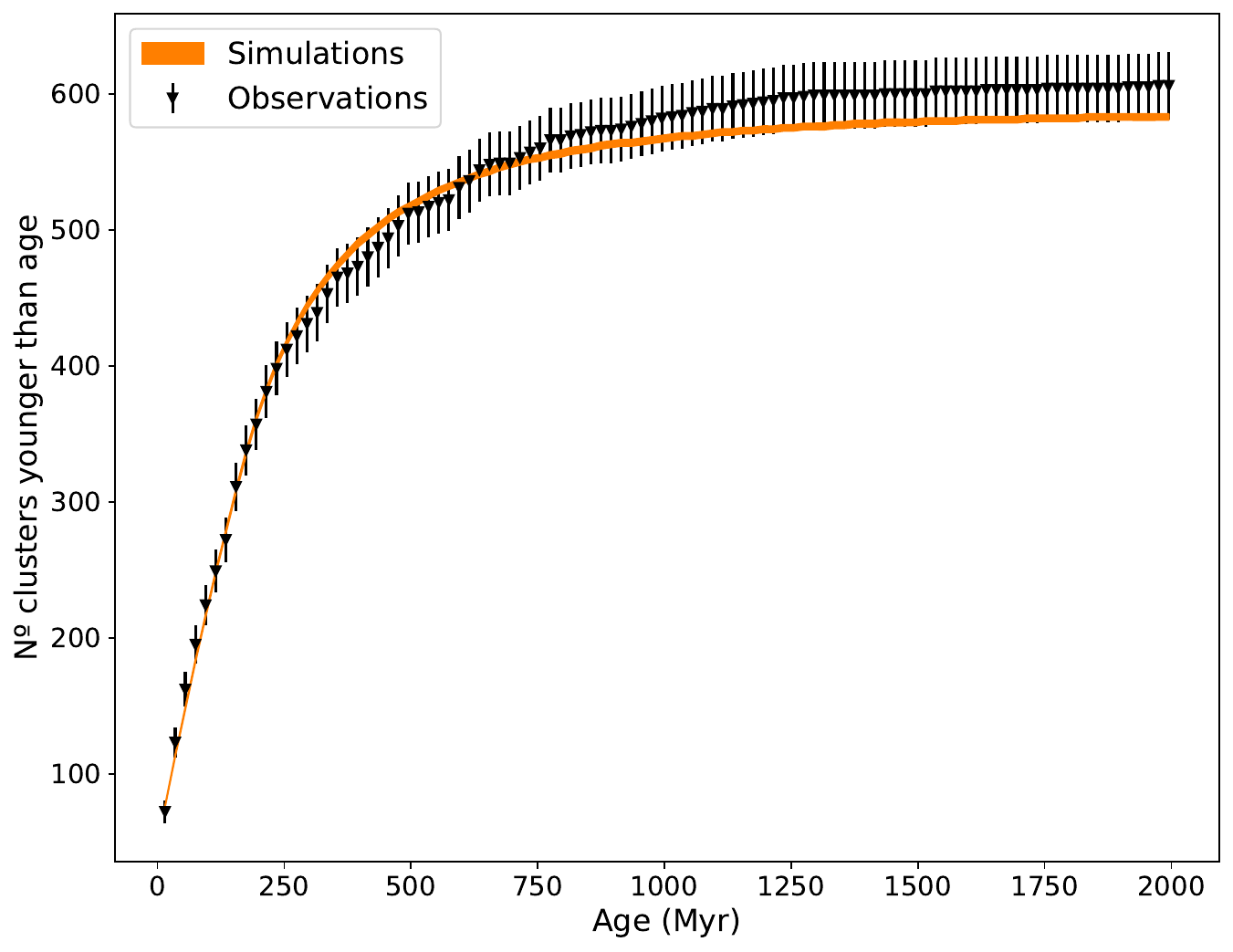}}
    \caption{Comparison of the evolution of the total number of OCs younger than time using $B_{SH} = 92$ pc, $D_{SH} = 50$ pc, $p_0 = 0.004$, and $t_0 = 11$ Myr. The error bars of the observations correspond to 1$\sigma$ Poisson errors. The values presented for the simulations are the average number observed in the ten independent runs and the thickness of the curve is given by the standard deviations.}
    \label{Nevol}
\end{figure}

Figure \ref{Nevol} shows the comparison of the evolution of the total number of clusters younger than a certain age. The good agreement shows that the model can accurately reproduce not only the SH evolution but also the timescale of disruption. In older versions of the model, this was not possible because we solely considered disruption due to encounters with GMCs. It was quickly realised that, to match the total number of OCs that survive at each age, the SH evolution would have to be compromised. The necessary increase in encounter probability would lead to an overestimated SH evolution. The introduction of the ICMF and the possibility for OCs to gradually lose mass, irrespective of their heights, enabled the model to reproduce the total number of surviving OCs without compromising the SH evolution.

Although ages $> 1$ Gyr were not used to constrain the evolution of the total number of clusters, the model shows good agreement for older ages. Furthermore, it is possible to find combinations of parameters that improve the adjustment for ages $> 1$ Gyr. However, as we discussed above, the origin of the difference in completeness between the older age group and the younger ones is not clear. It is interesting, however, that the model under-predicts the total number of OCs for the older ages. This was expected, since the OCs were generated following the completeness trend of the younger age groups, which rapidly decay with heliocentric distance, in contrast to the older group, for which this decay is much less accentuated. 

It is important to note that the incompleteness of the sample can impact this adjustment. Due to the cumulative nature of the distribution, it is likely that we are progressively missing more OCs as we consider older ages, raising the possibility that total intensity of disruption is being overestimated. Thus, it is possible that $t_0$ is higher than the one estimated from the fits using our model. Additionally, the parameters related to the encounters with GMCs might also be affected due to their indirect dependence on the overall disruption caused by stellar and dynamical evolution, as was discussed previously.

\subsection{{Other mechanisms that could affect the scale height evolution}}
{
In the previous sections, we have shown that our model, with realistic parameters, can largely explain the observations of increasing scale height with age. Still, it is important to assess the impact of other processes that may potentially cause similar thickening effects; namely, vertical heating from GMCs and spiral arms, and radial migration.}

{We begin by noting that while vertical heating and radial migration are well established in the dynamics of single stars, they are not established mechanisms of altering the vertical scale distribution of OCs. This is primarily because OCs are often destroyed by the interactions driving these phenomena.
}

\paragraph{Vertical heating:}
{From their simulations, \citet{gustafsson_gravitational_2016} note that:
(1) Spiral arms primarily contribute to in-plane heating but minimally affect the vertical motions of OCs. 
(2) Interactions with GMCs are the dominant driver of vertical heating. However, they also observe that these interactions limit vertical heating in OCs due to the high likelihood of OC destruction before encounters allow the OCs to reach high altitudes. Only a small fraction of massive clusters (typically 0.5\%), which are more resistant to destruction, reach high altitudes. Additionally, clusters that are not scattered to higher altitudes experience faster destruction. Thus, for the majority of OCs, which have considerably lower masses, vertical scattering is limited due to rapid destruction.
}

{Furthermore, the kinematic study by \citet{Tarricq_2021} shows no significant evidence for OCs reaching higher altitudes due to vertical heating. They find that the heating rate is found to be similar to stars in the GP but not perpendicular to the plane. They interpret this as a result of clusters not reaching high altitudes and older ages because they are disrupted before, and/or that GMCs, which are the primary drivers of vertical scattering in field stars, are less effective at scattering OCs, or their primary effect is OC destruction. Current observational studies therefore do not support significant vertical heating of the OC population by GMCs and spiral arms in the solar neighbourhood.

\paragraph{Radial migration}
{To our knowledge, the effect of radial migration on the vertical distribution of OCs has not been the object of dynamical modelling. For stars for which detailed modeling exists, the effect of radial migration on disc thickening remains debated, with some studies arguing that it does little for Galactic disc thickening \citep[e.g.][]{2012A&A...548A.127M}. 
}

{In their study of orbit migration and heating using red clump stars, \citet{2020ApJ...896...15F} discuss the in-plane birth radius of the Sun, indicating a migration distance of approximately 2 kpc for 4.6 Gyr old stars. This corresponds to a displacement of less than 1 kpc over the time frame of our simulations. \citet{2022MNRAS.511.5639L} also find radial migration to be inefficient over short timescales. They find an  average migration of $\sim 1.4$ kpc for a 2 Gyr population at 7-9 kpc from the centre of the Galaxy. The value would be even lower for younger populations.
}

{Observational studies of OCs also suggest limited radial migration. \citet{2020MNRAS.495.2673C} report that most clusters migrate less than 1 kpc/Gyr. \citet{2022MNRAS.509..421N} estimate a migration rate of 1 kpc/Gyr for objects up to 2 Gyr old. \citet{2023A&A...679A.122V} provide a detailed comparison of the responses of OCs and field stars to radial migration, finding that OCs younger than 2–3 Gyr are more resistant to perturbations than field stars, tend to move along quasi-circular orbits, and exhibit distinct differences in the time evolution of their kinematic and orbital properties compared to field stars.
}

{From these studies, we see that dynamical modelling and observations consistently support small migrations of $\sim$1 kpc over our simulation period. At this range, well below the disc scale length, changes in the galactic potential have minimal impact on plane-vertical coupling, implying negligible effects on SH evolution.
}

{As an additional consideration, following the discussion in Sect.~\ref{sec:p0}, the value of $p_0$ aligns well with other independent astrophysical constraints. This further supports our model, suggesting that alternative vertical heating mechanisms do not play a significant role in OC scale height evolution over the observed timescale.
}

\section{Conclusions and future outlook}

In this study, we propose a model to explain the increase in the SH with age of the OC population {in the solar neighbourhood}. We argue that the apparent thickening of the disc seen with OCs can be largely explained as a consequence of a stronger disruption of OCs near the GP, due to interactions with the disc; namely, encounters with GMCs. To test our hypothesis, we developed a computation model that forms OCs with different initial masses and follows their orbits while subjecting them to different disruption mechanisms. These mechanisms are: mass loss due to stellar and dynamical evolution, and encounters with GMCs in which the probability of the encounters depends on the heights of the OCs. We compared the simulations produced with our model to the \textit{Gaia}-based catalogue of OCs by \citet{DiasCat}. We find that:

\begin{itemize}
    \item The number of observed OCs decreases with heliocentric distance, indicating that the sample is incomplete. The sample incompleteness was taken into account in our model set-up.
    
    \item With optimal parameters, the model remarkably reproduces the SH evolution from the observations up to 1 Gyr. For the OCs with $1000 <$ age (Myr) $< 2000$, the predicted SH from the model starts to deviate from the observations. Nonetheless, it remains within the error bars of the observations. The deviation can be associated with effects of incompleteness and/or the simplicity of the model. 
    
    \item The model successfully reproduces the total number of OCs that survive with age. This was achieved by generating clusters following an ICMF, as well as the implementation of disruption caused by mass loss from stellar and dynamical evolution, which gradually dissolve the OCs independently of their heights.


    \item {The inferred dissolution timescale, $t_0 = 11$ Myr, is consistent with values obtained from N-body simulations \citep[approximately 9--18 Myr;][]{BaumgardtMakino,LamersGieles2006}.}
    
    \item {The inferred parameters related to GMCs show good agreement with previous estimates derived using different methods. Specifically, the dissolution scale height, $D_{SH} = 50$ pc, which measures the vertical extent of the destructive effects of OC encounters with GMCs, aligns well with the thickness of the GMC distribution and the general CO layer in the solar vicinity \citep{Stark&Lee2005, 1994ApJ...433..687M}. Furthermore, the encounter probability per Myr, $p_0 = 0.004$, enables an estimation of the average time for an object with an orbit similar to that of the Sun to encounter a GMC as approximately 700 Myr. This is consistent with the estimate for the Sun provided by \citet{2019MNRAS.489.5165K} using different approaches.}
    
    \item {Over the timescale and spatial volume covered by our simulations, the effects of vertical heating and radial migration on OCs can be considered negligible.}

\end{itemize}    

Although the model successfully reproduces the SH evolution and the total number of clusters that survive with age, there is still room for improvement. Future perspectives include:

\begin{itemize}
    \item {Analysing the kinematic evolution of the OC population. This requires one to account for initial velocities, rather than simply ‘drop’ the OCs as was done in this work. This approach is now feasible, given that \textit{Gaia} provides both proper motions and radial velocities for large numbers of clusters. We anticipate that the initial velocity dispersion of OCs will influence the inferred values of $B_{SH}$, $D_{SH}$, and the overall dynamics of the OC system.}

    \item The model considers the OCs to be completely disrupted in single encounters. Although this assumption might hold for OCs with small masses, it is possible that the more massive OCs survive single encounters. It can be interesting to explore how other intrinsic properties of the OCs such as the density and radius impact the results and, perhaps, explain the discrepancy of the predicted SH of OCs with ages $> 1$ Gyr. 

    \item Finally, the fluctuations of the observed SH in the first $150$ Myr (e.g. figure~\ref{degeneracy}) deserve a closer look. They likely encode the structure and kinematics of the distribution of OCs at birth.
\end{itemize}

\begin{acknowledgements}
This work was partially supported by the Portuguese Funda\c c\~ao para a Ci\^encia e a Tecnologia (FCT) through the Portuguese Strategic Programmes UIDB/FIS/00099/2020 and UIDP/FIS/00099/2020 for CENTRA. DA acknowledges support from the FCT/CENTRA PhD scholarship UI/BD/154465/2022.
\end{acknowledgements}

\bibliographystyle{aa}
\bibliography{biblio}
\end{document}